\documentclass{aa}
\usepackage{graphicx}
\usepackage{lscape}

\begin{document}

\title{The X-ray emission from Young Stellar Objects \\ in the $\rho$ Ophiuchi  cloud core as seen by XMM-Newton}
\subtitle{}

\author{H. Ozawa\inst{1} \and N. Grosso\inst{1} \and T. Montmerle\inst{1}}

\institute{Laboratoire d'Astrophysique de Grenoble, Universit\'e Joseph-Fourier, F-38041 Grenoble Cedex 9, France }
\date{}

\abstract{
We observed the main core F of the $\rho$ Ophiuchi  cloud, an active star-forming region located at $d \sim 140$ pc, using {\it XMM-Newton} with an exposure of 33 ks. We detect 87 X-ray sources within the 30$\arcmin$ diameter field-of-view of the {\it EPIC} imaging detector array. We cross-correlate the positions of {\it XMM-Newton} X-ray sources with previous X-ray and infrared (IR) catalogs: 25 previously unknown X-ray sources are found from our observation; 43 X-ray sources are detected by both {\it XMM-Newton} and {\it Chandra}; 68 {\it XMM-Newton} X-ray sources have 2MASS near-IR counterparts. We show that {\it XMM-Newton} and {\it Chandra} have comparable sensitivity for point source detection when the exposure time is set to $\sim$ 30 ks for both. We detect X-ray emission from 7 Class I sources, 26 Class II sources, and 17 Class III sources. The X-ray detection rate of Class I sources is very high (64 \%), which is consistent with previous {\it Chandra} observations in this area. We propose that 15 X-ray sources are new class III candidates, which doubles the number of known Class III sources, and helps to complete the census of YSOs in this area. We also detect X-ray emission from two young bona fide brown dwarfs, GY310 and GY141, out of three known in the field of view. GY141 appears brighter by nearly two orders of magnitude than in the {\it Chandra} observation. We extract X-ray light curves and spectra from these YSOs, and find some of them showed weak X-ray flares. We observed an X-ray flare from the bona fide brown dwarf GY310. We find as in the previous {\it Chandra} observation of this region that Class I sources tend to have higher temperatures and heavier X-ray absorptions than Class II and III sources.
}

\titlerunning{XMM-Newton observation of the $\rho$ Oph cloud}
\authorrunning{Ozawa et al.}
\maketitle

\section{Introduction}

Young low-mass stars are known to be ubiquitous X-ray emitters (Montmerle et al.\ 1993; Feigelson and Montmerle 1999). 
Their X-ray luminosities are $\sim 10^4-10^5$ times higher than that of the present-day Sun.
YSOs are classified from IR and sub-millimeter spectral energy distributions (Andr\'e \& Montmerle 1994) into four classes: Class 0 and Class I protostars, Class II (classical T Tauri stars) and Class III (weak-lined T Tauri stars) sources, interpreted as a time evolution sequence from Class 0 (age $\sim 10^4$ yrs) to Class III (age $\ga 10^6$ yrs).
It is generally accepted that enhanced magnetic activity of YSOs produces X-rays emitted by high temperature plasmas as a result of heating by magnetic reconnection events, which is called the solar paradigm because of the analogy with the X-ray emission of the Sun (e.g. as seen with {\it Yohkoh}).

The $\rho$ Ophiuchi cloud is a well-known nearby active star-forming region at $d \sim 140$ pc,
which has been observed over twenty years by practically all X-ray observatories: {\it Einstein} (Montmerle et al.\ 1983), {\it ROSAT} (Casanova 1994; Casanova et al.\ 1995;  Grosso et al.\ 1997; Grosso et al.\ 2000; Grosso 2001), {\it ASCA} (Koyama et al.\ 1994; Kamata et al.\ 1997; Tsuboi et al.\ 2000), and  {\it Chandra} (Imanishi et al.\ 2001a; Imanishi et al.\ 2001b; Imanishi et al.\ 2003).
To investigate the YSO X-ray emission in the dense CO core F of the $\rho$ Ophiuchi  cloud
 (Loren et al.\ 1990), we took with {\it XMM-Newton} a 33 ks exposure of this region, as part of the EPIC Guarantee Time program, and
 we report here the results of this observation.
In Sect. \ref{sec_obana}, we present observation and data analysis, including source detection,
  extraction of X-ray light curves and X-ray spectra from detected sources.
We compare our {\it XMM-Newton} results with previously obtained {\it Chandra} results in this region in Sect. \ref{sec_chan}. 
 We discuss the IR properties of detected sources in Sect. \ref{sec_ir}.
 We present the X-ray detections of young bona fide brown dwarfs in Sect. \ref{sec_bd},
 and the X-ray properties of Class I, II and III sources in Sect. \ref{sec_x}.
Finally, we summarize all results in Sect. \ref{sec_sum}.

\section{Observation and data analysis}\label{sec_obana}

The main (densest) core F of the $\rho$ Ophiuchi  cloud was observed 
with {\it XMM-Newton} for 33 ks on February 19, 2001
 pointing at $\alpha_{\rm{J2000}}$ $\rm{= 16^{h}27^{m}26.0^{s}}$ and $\delta_{\rm{J2000}}$ $\rm{=-24\degr40\arcmin48.0\arcsec}$.
{\it XMM-Newton} is the X-ray astronomical observatory of the European Space Agency 
which was launched on December 10, 1999.
{\it XMM-Newton} has three co-aligned X-ray telescopes with 6$\arcsec$-FWHM angular
 resolution at 1.5 keV and with large total effective area up to 5000 cm$^{-2}$ at 1 keV (Jansen et al.\ 2001). 
The {\it European Photon Imaging Camera} (EPIC), i.e. two MOS (namely, MOS1 and MOS2) and one PN X-ray CCD arrays, 
 are placed on each focal plane of the three X-ray telescopes.
EPIC provides imaging capability over a wide field-of-view of 30$\arcmin$ diameter, and 
spectroscopic capability  with moderate spectral resolution ($\sim$ 60 eV at 1 keV) 
in the 0.2$-$10.0 keV band. In this observation,
EPIC was operated with the medium optical blocking filter and in the Full-Frame mode.

We reduced the data using the {\it XMM-Newton} {\it Science Analysis Software} (SAS version 5.4).
The photon event lists were obtained by running {\tt emchain}  and {\tt epchain}.  
We rejected bad pixels, and selected single, double, triple, and quadruple pixel events for MOS, and 
 single and double pixel events for PN using {\tt evselect}.
We excluded the time intervals where the X-ray counts from whole field of view 
are extremely high, above 4.4 counts s$^{-1}$ for MOS1 and MOS2, and 20 counts s$^{-1}$ for PN, indicating a high level of irradiation by solar protons, and hence a high background level on EPIC.
The sum of the final good-time-intervals are 31 ks for MOS1, 32 ks for MOS2, and 29 ks for PN.

\subsection{Source detection}

\begin{figure*}[p]
\begin{center}
\resizebox{12.5cm}{!}{\includegraphics[angle=0, bb=28 175 566 626]{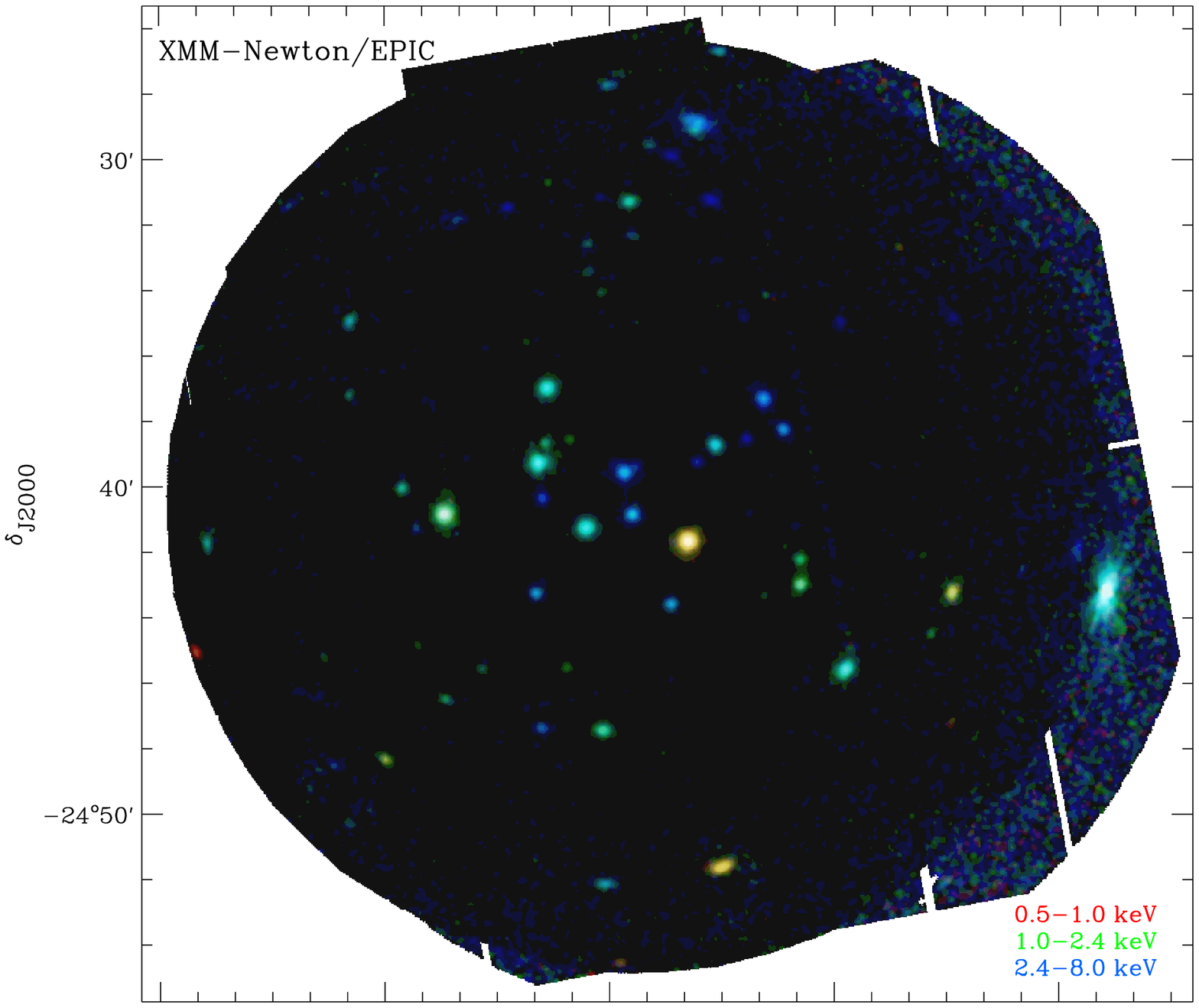}}\\
\resizebox{12.5cm}{!}{\includegraphics[angle=0]{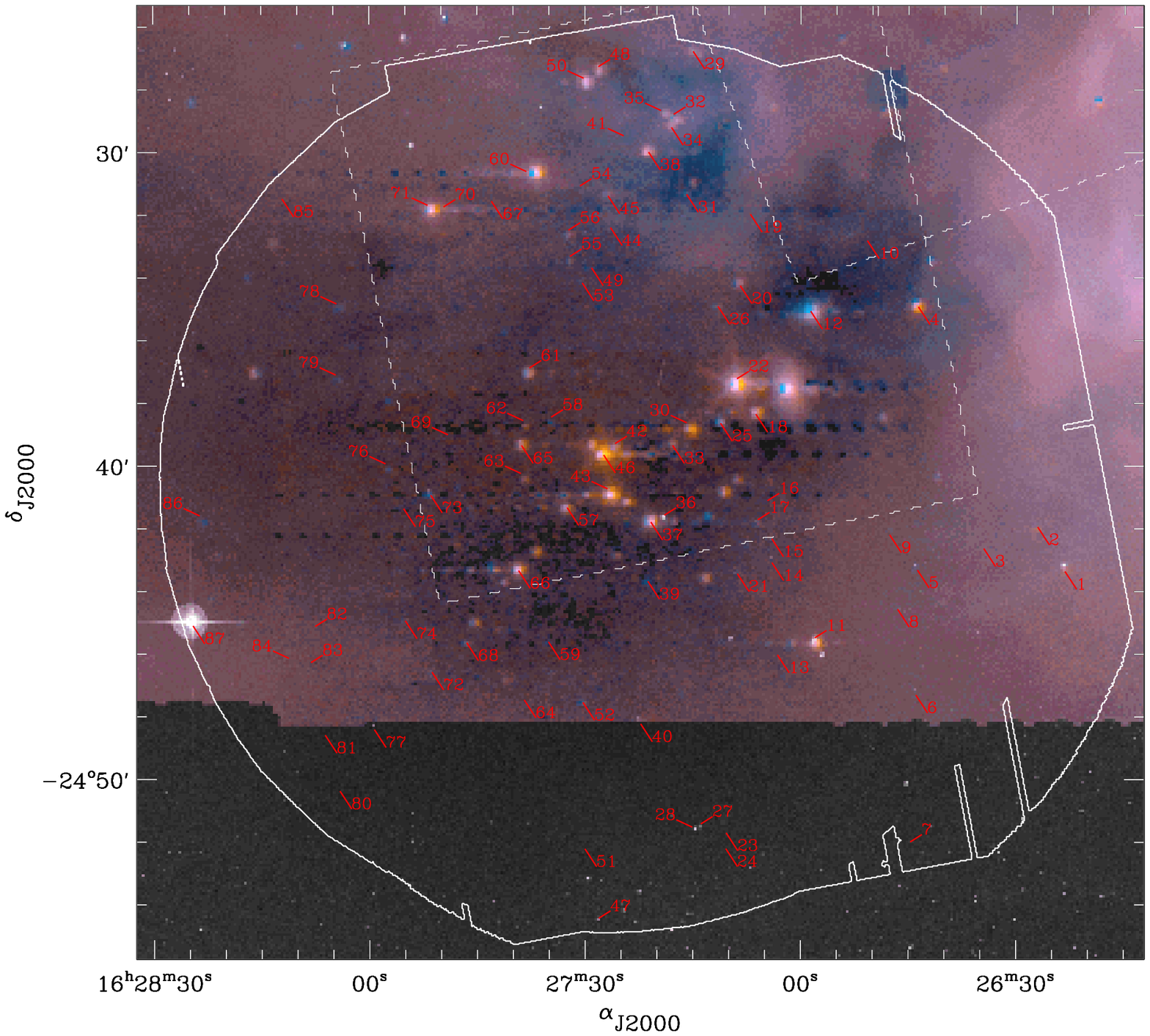}}
\caption{
{\it XMM-Newton} observation of the $\rho$ Ophiuchi cloud. 
The upper panel shows the tricolor image where EPIC MOS1, MOS2, and PN images are added together after vignetting correction. 
The image was smoothed with a 6\arcsec-FWHM Gaussian to increase the contrast.
The red, green, and blue colors code X-ray events in the 0.5$-$1.0 keV, 1.0$-$2.4 keV, and 2.4$-$8.0 keV band, respectively. 
The lower panel shows the same area with the positions of the 87 X-ray sources detected by {\it XMM-Newton} labeled by ROXN numbers (Table 1). The dashed lines show the field of view of {\it Chandra} (Imanishi et al. 2001a, Imanishi et al. 2003). The background image is a combination between DSS2-red (black \& white), ISOCAM LW2 6.7 $\mu$m (blue), and LW3 14.3 $\mu$m (red) filters.
}
\label{fig_tri}
\end{center}
\end{figure*}

Fig.~\ref{fig_tri} shows the resulting EPIC image where the MOS1, MOS2, and PN images were corrected for vignetting and co-added. 
Red, green, and blue colors code the X-ray events in the 0.5$-$1.0 keV, 1.0$-$2.4 keV, 2.4$-$8.0 keV energy bands, respectively. 
The image is smoothed to enhance the contrast.

We perform the source detection on the individual detector X-ray images, 
using the standard source detection method of SAS.
This method includes 4 steps, heritage of the {\it Einstein} and {\it ROSAT} source detection algorithms. 

1) The source candidates are obtained by the sliding cell method in {\tt eboxdetect}. 
   A small size cell surrounded by a larger size cell is moved over the X-ray image.
   The background counts are estimated from the area surrounding the small size cell.
   If the counts in the small size cell are significantly larger than in the background, 
   the cell is considered to have excess counts.
   After sliding the cell over the  whole region, 
   we obtain ``islands'' of excess counts,
   which are considered as source candidates. 

2) The local background image is created by {\tt esplinemap}.
   The source candidates obtained in step 1 are subtracted from the X-ray image,
   and a spline fit is applied to this ``swiss-cheese" image to obtain the background image.  

3) The sliding cell method is again performed with {\tt eboxdetect} to obtain more reliable source candidates. 
   This time, only the small size cell is moved over the X-ray image.
   The local background counts are estimated locally using the background image obtained in the previous step.

4) A two-dimensional fit is performed on the X-ray source candidates in {\tt emldetect} 
   using a model where the X-ray counts come from a point source,
   contamination by nearby point sources, and the background. 
   The point spread function is used as the model function of the point source.
   The local background created at step 2 is used as the background model.
   Through this task, the significance of the source candidates  is checked precisely 
   and source counts are obtained accurately.

This method can deal with images in different energy bands and different detectors simultaneously.
We use images in the 0.3$-$2.0 keV and 2.0$-$8.0 keV energy bands, 
and we deal with the MOS1, MOS2, and PN X-ray images simultaneously to obtain good statistics. 
For the areas covered by only two detectors or one detector, we can analyze only the images of the corresponding detector set. 

As a result of this protocol, we detect in total 87 X-ray sources with a likelihood threshold above 12, corresponding to the significance of $\sim$ 4.4 $\sigma$ for Gaussian statistics\footnote{There is a coding error in {\it emldetect} until SAS version 5.4 ({\it XMM-Newton}-News \#29, 11-Mar-2003), producing an overestimate of the source likelihood. We corrected the source likelihood using the table provided in the {\it XMM-Newton}-Newsletter.}, within the ~30$^\prime$ diameter field-of-view of the EPIC detectors. The positions of the detected 87 X-ray sources are indicated in the lower panel of Fig.~\ref{fig_tri}. 
Table 1 gives the coordinates, count rates, and the hardness ratios of the detected sources. For convenience, these sources are designated here ROXN-$n$ (for ``Rho Oph X-ray sources, Newton, number $n$")\footnote{The official naming convention for {\it XMM-Newton} sources is XMMU JHHMMSS.s$+/-$DDMMSS.}.
The hardness ratios are calculated after the formula $HR = (H - S)/(H + S)$,
 where $S$ and $H$ are counts in the 0.3$-$2.0 keV, and in the 2.0$-$8.0 keV energy band, respectively.   
\footnote{ROXN-1 was so bright that the CCD pixels on the source are strongly affected by pile-up. Hence $HR$ computed for this source is not reliable.}

The detected sources are cross-correlated with: X-ray source catalogs from {\it Einstein} (Montmerle et al.\ 1983), {\it ROSAT} (Casanova et al.\ 1995; Grosso et al.\ 2000), {\it ASCA} (Kamata el al.\ 1997), and {\it Chandra} (Imanishi et al.\ 2001a); near-IR source catalogs (e.g. Barsony et al.\ 1997; 2MASS = Two Micron All-Sky Survey); ISOCAM mid-IR catalog (Bontemps et al.\ 2001); and an optical catalog (Monet 1996). 
Table 2 shows the resulting cross-correlation:
 62 ROXN sources are identified with previously known X-ray sources, 
 hence we find 25 new X-ray sources from our observation, and
 68 ROXN sources have 2MASS near-IR counterparts.
Table 2 also lists the IR classifications from Bontemps et al.\ (2001), or Andr\'e \& Montmerle (1994).

As a result of these cross-identifications, we find 7 Class I sources, 26 Class II sources, and 17 Class III sources. We present 15 previously unknown Class III candidates in section \ref{sec_ir}.

\begin{table*}
\caption{
The list of the X-ray sources detected by {\it XMM-Newton}.
}
\tiny
\begin{tabular}{ccccrrrrrr}
\hline\hline
ROXN  & $\alpha_{\rm{J2000}}$ & $\delta_{\rm{J2000}}$  & err  & \multicolumn{2}{c}{MOS1} & \multicolumn{2}{c}{MOS2} & \multicolumn{2}{c}{PN}  \\
       &       &   & ($\arcsec$) & (cts/ks) & $H.R.$ & (cts/ks) & $H.R.$ &  (cts/ks) & $H.R.$  \\
(1) & (2) & (3) & (4) &  (5) & (6) & (7) & (8) & (9) & (10) \\
\hline
      1 &  16 26 23.7 & -24 43 16.2 &   0.1&   \multicolumn{1}{c}{-} &   \multicolumn{1}{c}{-} &   \multicolumn{1}{c}{-} &   \multicolumn{1}{c}{-} &              716.6(16.6)&              0.22(0.02) \\
      2 &  16 26 27.5 & -24 41 51.3 &   1.0&                6.7( 1.1)&              0.86(0.12) &                6.8( 1.1)&                     1.00&               11.7( 1.6)&              0.77(0.11) \\
      3 &  16 26 35.1 & -24 42 32.6 &   1.6&                0.8( 0.5)&              0.72(0.47) &                1.3( 0.6)&              0.73(0.38) &                4.8( 1.2)&              0.51(0.21) \\
      4 &  16 26 44.2 & -24 34 49.4 &   1.2&                2.1( 0.5)&              0.92(0.16) &                3.2( 0.7)&              0.66(0.20) &                3.3( 0.8)&              0.86(0.20) \\
      5 &  16 26 44.3 & -24 43 13.9 &   0.3&               18.4( 1.3)&           $-$0.62(0.06) &               16.2( 1.2)&           $-$0.71(0.06) &               45.7( 2.6)&           $-$0.69(0.04) \\
      6 &  16 26 44.5 & -24 47 13.4 &   1.2&                2.1( 0.5)&           $-$0.87(0.22) &                0.9( 0.4)&                  $-$1.00&                3.5( 0.7)&                  $-$1.00\\
      7 &  16 26 45.3 & -24 52 06.8 &   1.6&   \multicolumn{1}{c}{-} &   \multicolumn{1}{c}{-} &   \multicolumn{1}{c}{-} &   \multicolumn{1}{c}{-} &                7.8( 1.5)&           $-$0.04(0.19) \\
      8 &  16 26 47.1 & -24 44 28.9 &   1.5&                1.9( 0.5)&           $-$0.66(0.27) &                1.6( 1.1)&                  $-$1.00&                4.7( 0.9)&           $-$0.75(0.19) \\
      9 &  16 26 48.1 & -24 42 06.6 &   1.4&                1.1( 0.4)&           $-$0.15(0.37) &                0.7( 0.3)&           $-$0.16(0.49) &                2.2( 0.6)&                  $-$1.00\\
     10 &  16 26 51.3 & -24 32 42.7 &   1.0&                1.2( 0.3)&                  $-$1.00&                1.1( 0.4)&                  $-$1.00&                5.4( 0.8)&                  $-$1.00\\
     11 &  16 26 58.6 & -24 45 35.8 &   0.2&               40.2( 1.7)&              0.03(0.04) &               31.7( 1.6)&              0.01(0.05) &              111.4( 2.7)&              0.02(0.02) \\
     12 &  16 26 59.2 & -24 34 58.0 &   0.9&                2.1( 0.4)&                     1.00&                3.3( 0.6)&                     1.00&                4.9( 0.8)&                     1.00\\
     13 &  16 27 03.8 & -24 45 56.7 &   1.8&                0.5( 0.3)&           $-$0.11(0.59) &                0.7( 0.3)&              0.87(0.30) &                1.8( 0.6)&              0.49(0.32) \\
     14 &  16 27 04.6 & -24 42 59.9 &   0.3&               13.3( 0.9)&           $-$0.32(0.06) &               13.0( 0.8)&           $-$0.40(0.06) &               35.3( 1.5)&           $-$0.41(0.04) \\
     15 &  16 27 04.6 & -24 42 14.2 &   0.4&                4.2( 0.5)&           $-$0.47(0.11) &                6.3( 0.6)&           $-$0.26(0.09) &               14.3( 1.0)&           $-$0.41(0.07) \\
     16 &  16 27 05.2 & -24 41 13.2 &   1.8&                0.7( 0.2)&                     1.00&                0.7( 0.2)&                     1.00&                1.9( 0.5)&                     1.00\\
     17 &  16 27 06.5 & -24 41 49.7 &   1.3&                0.4( 0.2)&           $-$0.44(0.55) &                1.1( 0.3)&           $-$0.32(0.30) &                0.9( 0.4)&           $-$0.51(0.49) \\
     18 &  16 27 06.8 & -24 38 15.5 &   0.3&               11.0( 0.8)&              0.79(0.04) &               10.1( 0.7)&              0.72(0.05) &               30.4( 1.4)&              0.80(0.03) \\
     19 &  16 27 07.6 & -24 31 52.4 &   1.0&                0.6( 0.3)&              0.20(0.57) &                0.4( 0.3)&                     1.00&                2.9( 0.7)&           $-$0.79(0.24) \\
     20 &  16 27 09.0 & -24 34 09.3 &   1.3&                2.7( 0.6)&           $-$0.27(0.20) &                2.2( 0.9)&           $-$0.64(0.38) &                2.2( 0.6)&              0.33(0.24) \\
     21 &  16 27 09.4 & -24 43 21.2 &   1.4&                1.3( 0.3)&           $-$0.70(0.24) &                0.6( 0.3)&           $-$0.18(0.46) &                2.1( 0.5)&              0.22(0.25) \\
     22 &  16 27 09.4 & -24 37 19.4 &   0.2&               23.9( 1.1)&              0.90(0.02) &               23.6( 1.1)&              0.83(0.03) &               54.4( 1.8)&              0.87(0.02) \\
     23 &  16 27 10.9 & -24 51 37.8 &   1.6&                0.9( 0.4)&              0.87(0.32) &                1.6( 0.5)&              0.19(0.31) &                1.9( 0.8)&           $-$0.45(0.47) \\
     24 &  16 27 11.0 & -24 52 08.0 &   1.4&                1.6( 0.5)&              0.55(0.31) &                1.0( 0.4)&              0.22(0.41) &                1.0( 0.8)&              0.77(0.72) \\
     25 &  16 27 11.7 & -24 38 32.8 &   0.5&                3.0( 0.4)&                     1.00&                3.3( 0.5)&              0.91(0.09) &                8.2( 0.8)&              0.92(0.07) \\
     26 &  16 27 12.1 & -24 34 48.8 &   1.2&                1.2( 0.3)&                     1.00&                1.5( 0.3)&                     1.00&                2.1( 0.6)&              0.85(0.20) \\
     27 &  16 27 14.4 & -24 51 32.5 &   0.6&                4.5( 1.2)&           $-$0.84(0.18) &                6.0( 1.1)&           $-$0.82(0.12) &               22.0( 2.3)&           $-$0.83(0.08) \\
     28 &  16 27 15.1 & -24 51 38.7 &   0.1&               21.9( 1.5)&           $-$0.78(0.04) &               18.5( 1.3)&           $-$0.82(0.05) &               63.8( 2.8)&           $-$0.87(0.03) \\
     29 &  16 27 15.5 & -24 26 40.0 &   0.9&               10.4( 1.3)&              0.40(0.11) &               12.4( 1.4)&              0.46(0.10) &   \multicolumn{1}{c}{-} &   \multicolumn{1}{c}{-} \\
     30 &  16 27 15.9 & -24 38 43.6 &   0.2&               17.4( 0.9)&              0.40(0.05) &               19.5( 1.0)&              0.39(0.04) &               44.7( 1.6)&              0.31(0.03) \\
     31 &  16 27 16.4 & -24 31 14.2 &   0.7&                7.2( 0.8)&                     1.00&                6.0( 0.8)&              0.94(0.07) &               10.5( 1.1)&              0.96(0.05) \\
     32 &  16 27 18.2 & -24 28 52.7 &   0.2&               29.7( 2.2)&              0.91(0.03) &               34.3( 2.2)&              0.83(0.04) &               67.9( 3.2)&              0.92(0.03) \\
     33 &  16 27 18.4 & -24 39 15.4 &   0.5&                3.2( 0.4)&              0.84(0.08) &                4.4( 0.5)&              0.89(0.06) &               13.7( 1.7)&                     1.00\\
     34 &  16 27 18.6 & -24 29 05.9 &   0.6&                8.8( 1.1)&              0.20(0.12) &               11.2( 1.3)&              0.40(0.10) &               19.2( 1.9)&              0.48(0.09) \\
     35 &  16 27 19.4 & -24 28 45.4 &   0.8&                7.3( 1.2)&              0.88(0.09) &                7.0( 1.2)&              0.88(0.11) &               18.5( 2.1)&              0.53(0.09) \\
     36 &  16 27 19.5 & -24 41 40.9 &   0.1&              120.7( 2.2)&           $-$0.75(0.01) &              112.6( 2.1)&           $-$0.77(0.01) &              406.6( 4.4)&           $-$0.81(0.01) \\
     37 &  16 27 21.4 & -24 41 42.3 &   0.5&                3.4( 0.4)&           $-$0.56(0.11) &                2.4( 0.4)&           $-$0.84(0.14) &               11.6( 0.9)&           $-$0.59(0.07) \\
     38 &  16 27 21.7 & -24 29 53.2 &   0.9&                4.2( 0.7)&              0.83(0.12) &                3.0( 0.6)&                     1.00&                8.8( 1.2)&                     1.00\\
     39 &  16 27 21.8 & -24 43 35.8 &   0.3&               10.6( 0.7)&              0.57(0.05) &               10.5( 0.7)&              0.63(0.05) &               27.5( 1.3)&              0.67(0.04) \\
     40 &  16 27 22.8 & -24 48 09.0 &   1.3&                0.9( 0.3)&              0.15(0.35) &                1.2( 0.3)&              0.03(0.27) &                2.5( 0.6)&           $-$0.89(0.20) \\
     41 &  16 27 24.6 & -24 29 33.9 &   1.1&                1.2( 0.5)&           $-$0.12(0.40) &                3.4( 0.7)&              0.59(0.15) &                5.7( 1.1)&              0.56(0.17) \\
     42 &  16 27 26.4 & -24 39 23.7 &   1.0&                1.3( 0.4)&              0.67(0.21) &                1.9( 0.4)&                     1.00&                6.0( 0.8)&              0.65(0.10) \\
     43 &  16 27 26.9 & -24 40 50.4 &   0.2&               20.5( 0.9)&              0.78(0.03) &               22.0( 0.9)&              0.73(0.03) &               53.8( 1.6)&              0.73(0.02) \\
     44 &  16 27 27.0 & -24 32 18.0 &   0.9&                1.6( 0.4)&              0.34(0.24) &                3.0( 0.5)&              0.25(0.16) &                6.3( 0.9)&              0.62(0.12) \\
     45 &  16 27 27.4 & -24 31 17.2 &   0.4&               10.0( 0.9)&              0.19(0.09) &               12.5( 1.0)&           $-$0.03(0.08) &               29.8( 1.7)&              0.01(0.06) \\
     46 &  16 27 28.0 & -24 39 33.8 &   0.2&               28.9( 1.1)&              0.85(0.02) &               30.4( 1.1)&              0.81(0.02) &               75.5( 1.9)&              0.80(0.02) \\
     47 &  16 27 28.7 & -24 54 33.2 &   1.1&                2.8( 0.7)&           $-$0.54(0.23) &                3.5( 0.7)&                  $-$1.00&                8.4( 1.5)&           $-$0.76(0.19) \\
     48 &  16 27 28.7 & -24 27 21.4 &   2.4&                2.0( 0.6)&              0.64(0.30) &                2.5( 0.8)&           $-$0.30(0.33) &   \multicolumn{1}{c}{-} &   \multicolumn{1}{c}{-} \\
     49 &  16 27 29.7 & -24 33 35.0 &   1.3&                0.5( 0.2)&                     1.00&                1.2( 0.3)&              0.38(0.26) &                1.9( 0.5)&              0.71(0.24) \\
     50 &  16 27 30.2 & -24 27 43.4 &   0.9&                7.3( 0.9)&              0.44(0.11) &               10.9( 1.2)&              0.51(0.09) &   \multicolumn{1}{c}{-} &   \multicolumn{1}{c}{-} \\
     51 &  16 27 30.6 & -24 52 08.4 &   0.5&                7.4( 0.9)&              0.18(0.12) &                7.5( 0.8)&              0.22(0.11) &               23.6( 1.8)&              0.30(0.07) \\
     52 &  16 27 30.9 & -24 47 27.7 &   0.3&               13.4( 0.9)&           $-$0.10(0.07) &               13.1( 0.8)&           $-$0.11(0.07) &               35.8( 1.9)&           $-$0.10(0.05) \\
     53 &  16 27 31.0 & -24 34 03.9 &   0.9&                1.7( 0.3)&           $-$0.10(0.19) &                1.2( 0.4)&           $-$0.34(0.33) &                3.9( 0.7)&           $-$0.45(0.17) \\
     54 &  16 27 31.3 & -24 31 10.1 &   1.2&                1.0( 0.4)&                     1.00&                1.4( 0.4)&              0.88(0.21) &                3.2( 0.7)&                     1.00\\
     55 &  16 27 32.7 & -24 33 24.9 &   1.1&                1.4( 0.4)&              0.67(0.24) &                2.2( 0.5)&              0.28(0.20) &                3.7( 0.7)&              0.02(0.19) \\
     56 &  16 27 32.9 & -24 32 34.6 &   1.1&                2.3( 0.5)&              0.59(0.17) &                1.8( 0.7)&              0.75(0.32) &                4.3( 0.8)&              0.10(0.17) \\
     57 &  16 27 33.1 & -24 41 14.3 &   0.1&               44.8( 1.3)&              0.20(0.03) &               44.7( 1.3)&              0.18(0.03) &              120.0( 2.4)&              0.20(0.02) \\
     58 &  16 27 35.3 & -24 38 33.0 &   0.7&                2.1( 0.3)&           $-$0.59(0.16) &                1.7( 0.3)&           $-$0.89(0.14) &                4.0( 0.6)&           $-$0.80(0.17) \\
     59 &  16 27 35.7 & -24 45 31.7 &   0.8&                1.8( 0.3)&           $-$0.76(0.18) &                1.5( 0.3)&           $-$0.74(0.17) &                4.7( 0.7)&           $-$0.82(0.13) \\
     60 &  16 27 38.2 & -24 30 42.2 &   1.4&                0.4( 0.3)&              0.18(0.78) &                1.0( 0.3)&                  $-$1.00&                3.6( 0.8)&              0.04(0.23) \\
     61 &  16 27 38.3 & -24 36 58.6 &   0.1&               37.4( 1.3)&              0.12(0.04) &               38.0( 1.4)&              0.08(0.04) &               97.1( 2.4)&              0.09(0.02) \\
     62 &  16 27 38.6 & -24 38 39.6 &   0.5&                4.8( 0.5)&           $-$0.13(0.10) &                5.4( 0.5)&              0.00(0.09) &               14.1( 1.3)&           $-$0.10(0.09) \\
     63 &  16 27 38.9 & -24 40 20.9 &   0.6&                5.6( 0.5)&              0.98(0.03) &                6.0( 0.6)&              0.95(0.05) &               15.6( 3.2)&                     1.00\\
     64 &  16 27 39.1 & -24 47 23.1 &   0.5&                5.5( 0.6)&              0.78(0.07) &                4.8( 0.5)&              0.59(0.09) &               12.8( 1.1)&              0.71(0.06) \\
     65 &  16 27 39.4 & -24 39 15.4 &   0.1&               67.4( 1.7)&              0.13(0.03) &               68.9( 1.7)&              0.10(0.03) &              225.3(11.7)&              0.06(0.05) \\
     66 &  16 27 39.8 & -24 43 15.1 &   0.3&                9.6( 0.7)&              0.66(0.05) &                9.9( 0.7)&              0.68(0.05) &               25.6( 1.3)&              0.64(0.04) \\
     67 &  16 27 43.6 & -24 31 27.4 &   0.7&                3.1( 0.5)&              0.77(0.13) &                2.2( 0.5)&              0.79(0.16) &               10.3( 1.1)&                     1.00\\
     68 &  16 27 47.1 & -24 45 33.5 &   1.3&                1.7( 0.5)&           $-$0.30(0.32) &                1.0( 0.6)&           $-$0.18(0.55) &                3.9( 0.7)&              0.17(0.17) \\
     69 &  16 27 49.1 & -24 39 06.7 &   1.7&                0.8( 0.2)&              0.70(0.28) &                0.8( 0.2)&              0.38(0.31) &                1.0( 0.4)&              0.82(0.33) \\
     70 &  16 27 50.3 & -24 31 50.0 &   1.2&                2.3( 0.5)&              0.46(0.18) &                2.4( 0.5)&              0.64(0.18) &                6.5( 1.1)&              0.71(0.14) \\
     71 &  16 27 51.9 & -24 31 46.9 &   1.2&                1.0( 0.3)&                     1.00&                0.7( 0.3)&              0.47(0.47) &                5.5( 1.1)&              0.33(0.20) \\
     72 &  16 27 51.9 & -24 46 29.7 &   0.6&                4.0( 0.5)&           $-$0.24(0.13) &                3.2( 0.5)&           $-$0.20(0.14) &                8.1( 1.0)&              0.01(0.12) \\
     73 &  16 27 52.1 & -24 40 50.1 &   0.1&               50.5( 1.8)&           $-$0.27(0.03) &               54.9( 1.6)&           $-$0.25(0.03) &              171.2( 3.3)&           $-$0.37(0.02) \\
     74 &  16 27 55.6 & -24 44 51.3 &   1.1&                1.1( 0.3)&           $-$0.44(0.29) &                0.8( 0.3)&              0.13(0.38) &                2.5( 0.5)&                  $-$1.00\\
     75 &  16 27 55.8 & -24 41 16.5 &   0.8&                2.9( 0.5)&              0.74(0.13) &                1.6( 0.3)&              0.70(0.17) &                4.5( 0.7)&              0.68(0.14) \\
     76 &  16 27 57.8 & -24 40 02.3 &   0.6&                7.0( 0.8)&              0.22(0.11) &                7.7( 0.7)&           $-$0.03(0.09) &               19.6( 2.4)&              0.06(0.12) \\
     77 &  16 28 00.0 & -24 48 19.8 &   0.5&                6.3( 0.7)&           $-$0.71(0.09) &                5.1( 0.7)&           $-$0.67(0.11) &               18.4( 1.5)&           $-$0.54(0.08) \\
     78 &  16 28 04.6 & -24 34 55.7 &   0.4&                8.8( 0.8)&              0.33(0.09) &                5.7( 0.7)&              0.45(0.10) &               19.8( 1.6)&              0.22(0.08) \\
     79 &  16 28 04.7 & -24 37 11.9 &   0.9&                2.4( 0.5)&              0.09(0.22) &                1.8( 0.4)&           $-$0.27(0.23) &                6.9( 0.9)&              0.20(0.13) \\
     80 &  16 28 04.8 & -24 50 17.6 &   1.2&                1.0( 0.4)&           $-$0.29(0.42) &                1.8( 0.5)&              0.14(0.29) &                6.0( 1.2)&              0.20(0.19) \\
     81 &  16 28 06.8 & -24 48 30.3 &   0.8&                2.1( 0.5)&                     1.00&                1.5( 0.4)&              0.88(0.19) &                7.7( 1.2)&              0.76(0.13) \\
     82 &  16 28 08.2 & -24 45 13.8 &   1.1&                0.7( 0.3)&           $-$0.93(0.27) &                1.8( 0.4)&           $-$0.21(0.23) &                3.5( 0.8)&           $-$0.19(0.24) \\
     83 &  16 28 08.8 & -24 46 23.0 &   1.7&                0.7( 0.3)&                     1.00&                0.8( 0.4)&              0.31(0.43) &                3.3( 0.8)&                     1.00\\
     84 &  16 28 11.3 & -24 46 14.8 &   1.6&                0.9( 0.3)&                     1.00&                0.5( 0.3)&                     1.00&                2.9( 0.8)&              0.43(0.27) \\
     85 &  16 28 12.7 & -24 31 22.9 &   1.2&                4.6( 0.9)&                     1.00&                4.3( 0.8)&              0.56(0.15) &   \multicolumn{1}{c}{-} &   \multicolumn{1}{c}{-} \\
     86 &  16 28 23.7 & -24 41 41.5 &   0.9&                8.0( 0.9)&              0.16(0.11) &                9.9( 1.0)&              0.10(0.10) &   \multicolumn{1}{c}{-} &   \multicolumn{1}{c}{-} \\
     87 &  16 28 25.2 & -24 45 00.6 &   0.7&                7.1( 0.9)&                  $-$1.00&                8.7( 1.0)&                  $-$1.00&   \multicolumn{1}{c}{-} &   \multicolumn{1}{c}{-} \\
\hline
\end{tabular}
\scriptsize

(1) {\it XMM-Newton} source ID number, ``Rho Oph X-ray sources, Newton''.;
(5-10) The hyphens indicate that the X-ray sources are outside the field of view of the instrument.;
(6),(8),(10) The hardness ratio is calculated by the equation, $H.R. = (H - S) / (H + S)$, where $S$ and $H$ are counts in the 0.3$-$2.0 keV band and in the 2.0$-$8.0 keV band, respectively. Note that due to pile-up, $H.R.$  is maybe not reliable for source 1.
\label{xraytable}
\end{table*}

\begin{table*}

\caption{
Identification of the {\it XMM-Newton} sources with known X-ray sources and optical/infrared sources.
}
\tiny
\begin{tabular}{@{}c@{}ccccccccrrrrrcc@{}}
\hline\hline

\multicolumn{1}{c}{ROXN}  & E & R1 & R2 & A & RH & C & ISO & 2MASS J & \multicolumn{1}{c}{$J$} & \multicolumn{1}{c}{$H$} & \multicolumn{1}{c}{$K_\mathrm{s}$} & \multicolumn{1}{c}{$J-H$} & \multicolumn{1}{c}{$H-K_\mathrm{s}$} & NAME & Class \\
\multicolumn{1}{c}{(1)} & (2) & (3) & (4) &  (5) & (6) & (7) & (8) & (9) & \multicolumn{1}{c}{(10)} & \multicolumn{1}{c}{(11)} & \multicolumn{1}{c}{(12)} & \multicolumn{1}{c}{(13)} & \multicolumn{1}{c}{(14)} & (15) & (16) \\
\hline
 1\hspace{0.135cm} &\dotfill &  \dotfill &         - &  \dotfill &        F1 &         - &  \dotfill &  16262367-2443138 &     ~9.39 &     ~8.40 &     ~7.85&     ~0.99&     ~0.55 &                   DoAr25 &        II\\
           2$\ast$ &\dotfill &  \dotfill &         - &  \dotfill &  \dotfill &         - &  \dotfill &  16262753-2441535 &    ~14.04 &    ~11.42 &     ~9.98&     ~2.62&     ~1.44 &                     GY33 &        II\\
           3$\ast$ &\dotfill &  \dotfill &         - &  \dotfill &  \dotfill &         - &  \dotfill &          \dotfill & ~\dotfill & ~\dotfill & ~\dotfill&  \dotfill&  \dotfill &                 \dotfill &  \dotfill\\
 4\hspace{0.135cm} &\dotfill &  \dotfill &         - &  \dotfill &  \dotfill &         1 &        65 &  16264419-2434483 &    ~16.76 &    ~13.76 &    ~11.60&     ~3.00&     ~2.16 &                     WL12 &         I\\
 5\hspace{0.135cm} &\dotfill &        31 &         - &  \dotfill &        F2 &         - &        66 &  16264429-2443141 &    ~10.99 &    ~10.02 &     ~9.57&     ~0.97&     ~0.45 &                    GY112 &       III\\
           6$\ast$ &\dotfill &  \dotfill &         - &  \dotfill &  \dotfill &         - &  \dotfill &  16264441-2447138 &    ~11.83 &    ~11.00 &    ~10.63&     ~0.83&     ~0.37 &        BKLT162644-244711 &      nIII\\
           7$\ast$ &\dotfill &         - &  \dotfill &         - &  \dotfill &         - &         - &          \dotfill & ~\dotfill & ~\dotfill & ~\dotfill&  \dotfill&  \dotfill &                 \dotfill &  \dotfill\\
 8\hspace{0.135cm} &     C10 &  \dotfill &         - &  \dotfill &        F3 &         - &        69 &  16264705-2444298 &    ~12.33 &    ~11.12 &    ~10.56&     ~1.21&     ~0.56 &                    GY122 &       III\\
           9$\ast$ &\dotfill &  \dotfill &         - &  \dotfill &  \dotfill &         - &  \dotfill &  16264810-2442033 &    ~12.45 &    ~11.28 &    ~10.65&     ~1.17&     ~0.63 &                    GY125 &      nIII\\
          10$\ast$ &\dotfill &  \dotfill &         - &  \dotfill &  \dotfill &  \dotfill &  \dotfill &  16265128-2432419 &    ~15.30 &    ~14.47 &    ~13.89&     ~0.83&     ~0.58 &                    GY141 &        bd\\
11\hspace{0.135cm} &     C13 &        35 &         - &  \dotfill &        F7 &         - &        88 &  16265850-2445368 &     ~9.75 &     ~8.16 &     ~7.06&     ~1.59&     ~1.10 &                    SR24S &        II\\
12\hspace{0.135cm} &\dotfill &  \dotfill &         - &  \dotfill &  \dotfill &        10 &        90 &  16265916-2434588 &  $>$17.26 &  $>$17.82 &    ~15.13& ~\dotfill&   $>$2.69 &                     WL22 &        II\\
          13$\ast$ &\dotfill &  \dotfill &         - &  \dotfill &  \dotfill &         - &  \dotfill &          \dotfill & ~\dotfill & ~\dotfill & ~\dotfill&  \dotfill&  \dotfill &                 \dotfill &  \dotfill\\
14\hspace{0.135cm} &\dotfill &  \dotfill &         - &  \dotfill &        F8 &         - &        96 &  16270451-2442596 &    ~12.02 &    ~10.55 &     ~9.84&     ~1.47&     ~0.71 &                    GY193 &       III\\
15\hspace{0.135cm} &\dotfill &  \dotfill &         - &  \dotfill &        F9 &  \dotfill &        97 &  16270456-2442140 &    ~12.21 &    ~10.54 &     ~9.81&     ~1.67&     ~0.73 &                    GY194 &       III\\
16\hspace{0.135cm} &\dotfill &  \dotfill &         - &  \dotfill &  \dotfill &        15 &  \dotfill &          \dotfill & ~\dotfill & ~\dotfill & ~\dotfill&  \dotfill&  \dotfill &                 \dotfill &  \dotfill\\
17\hspace{0.135cm} &\dotfill &  \dotfill &         - &  \dotfill &  \dotfill &        17 &       102 &  16270659-2441488 &    ~12.43 &    ~11.40 &    ~10.77&     ~1.03&     ~0.63 &                    GY204 &        II\\
18\hspace{0.135cm} &\dotfill &  \dotfill &         - &        C5 &  \dotfill &        18 &       103 &  16270677-2438149 &  $>$17.28 &    ~14.30 &    ~10.97&   $>$2.98&     ~3.33 &                     WL17 &        II\\
          19$\ast$ &\dotfill &  \dotfill &         - &  \dotfill &  \dotfill &  \dotfill &  \dotfill &          \dotfill & ~\dotfill & ~\dotfill & ~\dotfill&  \dotfill&  \dotfill &                 \dotfill &  \dotfill\\
20\hspace{0.135cm} &\dotfill &        37 &         - &         7 &  \dotfill &        21 &       105 &  16270910-2434081 &    ~12.55 &    ~10.19 &     ~8.91&     ~2.36&     ~1.28 &                     WL10 &        II\\
          21$\ast$ &\dotfill &  \dotfill &         - &  \dotfill &  \dotfill &         - &  \dotfill &  16270931-2443196 &    ~14.30 &    ~12.21 &    ~11.06&     ~2.09&     ~1.15 &                    GY212 &      nIII\\
22\hspace{0.135cm} &\dotfill &  \dotfill &         - &         8 &  \dotfill &        23 &       108 &  16270943-2437187 &    ~16.79 &    ~11.05 &     ~7.14&     ~5.74&     ~3.91 &                     EL29 &         I\\
          23$\ast$ &\dotfill &         - &  \dotfill &         - &  \dotfill &         - &         - &          \dotfill & ~\dotfill & ~\dotfill & ~\dotfill&  \dotfill&  \dotfill &                 \dotfill &  \dotfill\\
          24$\ast$ &\dotfill &         - &  \dotfill &         - &  \dotfill &         - &         - &          \dotfill & ~\dotfill & ~\dotfill & ~\dotfill&  \dotfill&  \dotfill &                 \dotfill &  \dotfill\\
25\hspace{0.135cm} &\dotfill &  \dotfill &         - &  \dotfill &  \dotfill &        25 &       114 &  16271171-2438320 &  $>$18.60 &    ~15.06 &    ~11.06&   $>$3.54&     ~4.00 &                     WL19 &       III\\
          26$\ast$ &\dotfill &  \dotfill &         - &  \dotfill &  \dotfill &  \dotfill &       115 &  16271213-2434491 &    ~15.62 &    ~13.11 &    ~11.49&     ~2.51&     ~1.62 &                     WL11 &        II\\
27\hspace{0.135cm} &      20 &         - &        22 &         - &  \dotfill &         - &         - &  16271448-2451334 &    ~11.45 &    ~10.69 &    ~10.38&     ~0.76&     ~0.31 &                  ROXs20A &       III\\
28\hspace{0.135cm} &      20 &         - &  \dotfill &         - &       F14 &         - &         - &  16271513-2451388 &    ~10.62 &     ~9.77 &     ~9.39&     ~0.85&     ~0.38 &                  ROXs20B &       III\\
29\hspace{0.135cm} &\dotfill &        C6 &         - &  \dotfill &  \dotfill &        26 &  \dotfill &  16271545-2426398 &    ~17.42 &    ~13.46 &    ~10.79&     ~3.96&     ~2.67 &                    IRS34 &        II\\
30\hspace{0.135cm} &     C16 &        39 &         - &  \dotfill &       F16 &        29 &       121 &  16271587-2438433 &    ~13.89 &    ~11.26 &     ~9.59&     ~2.63&     ~1.67 &                     WL20 &        II\\
31\hspace{0.135cm} &\dotfill &  \dotfill &         - &  \dotfill &  \dotfill &        30 &  \dotfill &  16271643-2431145 &  $>$18.70 &  $>$17.69 &    ~15.09& ~\dotfill&   $>$2.60 &           16271643-24311 &      nIII\\
32\hspace{0.135cm} &\dotfill &        40 &         - &  \dotfill &  \dotfill &        34 &       125 &  16271817-2428526 &  $>$17.87 &    ~14.69 &    ~10.56&   $>$3.18&     ~4.13 &                      WL5 &       III\\
33\hspace{0.135cm} &\dotfill &  \dotfill &         - &  \dotfill &  \dotfill &        35 &       127 &  16271838-2439146 &  $>$18.67 &    ~15.54 &    ~12.23&   $>$3.13&     ~3.31 &                    GY245 &        II\\
34\hspace{0.135cm} &\dotfill &  \dotfill &         - &        C6 &  \dotfill &        36 &       128 &  16271848-2429059 &    ~14.61 &    ~11.50 &     ~9.68&     ~3.11&     ~1.82 &                      WL4 &        II\\
35\hspace{0.135cm} &\dotfill &  \dotfill &         - &  \dotfill &  \dotfill &        38 &       129 &  16271921-2428438 &  $>$17.61 &    ~14.66 &    ~11.49&   $>$2.95&     ~3.17 &                      WL3 &        II\\
36\hspace{0.135cm} &      21 &        41 &         - &        9A &       F17 &        40 &       130 &  16271951-2441403 &     ~9.42 &     ~8.63 &     ~8.41&     ~0.79&     ~0.22 &                  SR12A-B &       III\\
37\hspace{0.135cm} &\dotfill &  \dotfill &         - &  \dotfill &  \dotfill &        43 &       132 &  16272146-2441430 &    ~15.22 &    ~11.25 &     ~8.48&     ~3.97&     ~2.77 &                    IRS42 &        II\\
38\hspace{0.135cm} &\dotfill &       C7? &         - &        10 &  \dotfill &        44 &       134 &  16272180-2429533 &  $>$18.65 &    ~15.38 &    ~10.83&   $>$3.27&     ~4.55 &                      WL6 &         I\\
39\hspace{0.135cm} &     C18 &  \dotfill &         - &        C7 &  \dotfill &         - &       133 &  16272183-2443356 &    ~17.27 &    ~13.20 &    ~10.78&     ~4.07&     ~2.42 &                    GY253 &       III\\
40\hspace{0.135cm} &\dotfill &  \dotfill &         - &  \dotfill &       F19 &         - &         - &  16272297-2448071 &    ~10.92 &     ~9.86 &     ~9.39&     ~1.06&     ~0.47 &                    WSB49 &        II\\
41\hspace{0.135cm} &\dotfill &  \dotfill &         - &  \dotfill &  \dotfill &        50 &  \dotfill &  16272463-2429353 &  $>$18.71 &    ~14.78 &    ~12.43&   $>$3.93&     ~2.35 &                    GY259 &      nIII\\
42\hspace{0.135cm} &\dotfill &  \dotfill &         - &  \dotfill &  \dotfill &        53 &  \dotfill &  16272648-2439230 &    ~15.69 &    ~12.07 &     ~9.95&     ~3.62&     ~2.12 &                    GY262 &        II\\
43\hspace{0.135cm} &\dotfill &        43 &         - &  \dotfill &       F21 &        54 &       141 &  16272693-2440508 &  $>$18.53 &    ~13.52 &     ~9.74&   $>$5.01&     ~3.78 &              IRS43/YLW15 &         I\\
44\hspace{0.135cm} &\dotfill &  \dotfill &         - &  \dotfill &  \dotfill &        55 &  \dotfill &  16272706-2432175 &  $>$17.81 &    ~14.51 &    ~12.42&   $>$3.30&     ~2.09 &                    GY266 &      nIII\\
45\hspace{0.135cm} &\dotfill &        44 &         - &  \dotfill &       F22 &        56 &       142 &  16272738-2431165 &    ~12.35 &    ~10.38 &     ~9.32&     ~1.97&     ~1.06 &                   VSSG25 &        II\\
46\hspace{0.135cm} & \dotfill$\dagger$ &  \dotfill &         - &        9B &  \dotfill &        57 &       143 &  16272802-2439335 &  $>$16.56 &    ~13.68 &    ~10.38&   $>$2.88&     ~3.30 &             IRS44/YLW16A &         I\\
47\hspace{0.135cm} &\dotfill &         - &  \dotfill &         - &       F24 &         - &         - &  16272873-2454317 &    ~12.60 &    ~12.02 &    ~11.69&     ~0.58&     ~0.33 &   0600-20531406$\ddagger$ &  \dotfill\\
48\hspace{0.135cm} &\dotfill &  \dotfill &         - &  \dotfill &  \dotfill &        59 &       144 &  16272844-2427210 &    ~15.74 &    ~12.31 &    ~10.10&     ~3.43&     ~2.21 &                    IRS45 &        II\\
49\hspace{0.135cm} &\dotfill &  \dotfill &         - &  \dotfill &  \dotfill &        61 &       146 &  16272996-2433365 &  $>$18.70 &    ~15.58 &    ~12.43&   $>$3.12&     ~3.15 &        BKLT162730-243336 &      nIII\\
50\hspace{0.135cm} &\dotfill &  \dotfill &         - &  \dotfill &  \dotfill &        63 &       147 &  16273018-2427433 &    ~15.32 &    ~11.52 &     ~9.02&     ~3.80&     ~2.50 &                    IRS47 &        II\\
          51$\ast$ &\dotfill &         - &  \dotfill &         - &  \dotfill &         - &         - &          \dotfill & ~\dotfill & ~\dotfill & ~\dotfill&  \dotfill&  \dotfill &                 \dotfill &  \dotfill\\
52\hspace{0.135cm} &\dotfill &        46 &         - &  \dotfill &       F26 &         - &       149 &  16273084-2447268 &    ~12.24 &    ~10.42 &     ~9.50&     ~1.82&     ~0.92 &        BKLT162730-244726 &       III\\
53\hspace{0.135cm} &\dotfill &        47 &         - &  \dotfill &  \dotfill &        64 &       148 &  16273105-2434032 &    ~13.43 &    ~11.36 &    ~10.39&     ~2.07&     ~0.97 &                    GY283 &      nIII\\
54\hspace{0.135cm} &\dotfill &  \dotfill &         - &  \dotfill &  \dotfill &        65 &  \dotfill &          \dotfill & ~\dotfill & ~\dotfill & ~\dotfill&  \dotfill&  \dotfill &                 \dotfill &  \dotfill\\
55\hspace{0.135cm} &\dotfill &  \dotfill &         - &  \dotfill &  \dotfill &        66 &       152 &  16273267-2433239 &    ~16.15 &    ~12.74 &    ~10.90&     ~3.41&     ~1.84 &                    GY289 &       III\\
56\hspace{0.135cm} &\dotfill &        48 &         - &  \dotfill &  \dotfill &        67 &       154 &  16273285-2432348 &    ~16.19 &    ~12.74 &    ~10.96&     ~3.45&     ~1.78 &                    GY291 &        II\\
57\hspace{0.135cm} &\dotfill &  \dotfill &         - &  \dotfill &       F27 &        69 &       155 &  16273311-2441152 &    ~11.32 &     ~9.13 &     ~7.81&     ~2.19&     ~1.32 &                    GY292 &        II\\
58\hspace{0.135cm} &\dotfill &  \dotfill &         - &  \dotfill &  \dotfill &        70 &       156 &  16273526-2438334 &    ~11.28 &    ~10.23 &     ~9.67&     ~1.05&     ~0.56 &                    GY295 &      nIII\\
59\hspace{0.135cm} &\dotfill &  \dotfill &         - &  \dotfill &       F28 &         - &       157 &  16273566-2445325 &    ~12.71 &    ~11.47 &    ~10.88&     ~1.24&     ~0.59 &                    GY296 &       III\\
60\hspace{0.135cm} &\dotfill &       C10 &         - &  \dotfill &  \dotfill &        74 &  \dotfill &  16273812-2430429 &    ~12.54 &    ~10.54 &     ~9.66&     ~2.00&     ~0.88 &                    IRS50 &       III\\
61\hspace{0.135cm} &     C20 &  \dotfill &         - &  \dotfill &       F29 &        75 &       163 &  16273832-2436585 &    ~11.38 &     ~9.43 &     ~8.27&     ~1.95&     ~1.16 &                    IRS49 &        II\\
62\hspace{0.135cm} &\dotfill &  \dotfill &         - &  \dotfill &  \dotfill &        76 &       164 &  16273863-2438391 &    ~13.27 &    ~11.93 &    ~11.08&     ~1.34&     ~0.85 &                    GY310 &    II, bd\\
63\hspace{0.135cm} &\dotfill &  \dotfill &         - &  \dotfill &  \dotfill &        77 &       165 &  16273894-2440206 &    ~16.54 &    ~13.91 &    ~12.29&     ~2.63&     ~1.62 &                    GY312 &        II\\
          64$\ast$ &\dotfill &  \dotfill &         - &         - &  \dotfill &         - &  \dotfill &          \dotfill & ~\dotfill & ~\dotfill & ~\dotfill&  \dotfill&  \dotfill &                 \dotfill &  \dotfill\\
65\hspace{0.135cm} &\dotfill &        51 &         - &  \dotfill &       F30 &        78 &       166 &  16273942-2439155 &    ~10.75 &     ~9.21 &     ~8.46&     ~1.54&     ~0.75 &                    GY314 &        II\\
66\hspace{0.135cm} &\dotfill &  \dotfill &         - &  \dotfill &  \dotfill &        79 &       167 &  16273982-2443150 &    ~17.05 &    ~12.13 &     ~8.99&     ~4.92&     ~3.14 &                    IRS51 &         I\\
67\hspace{0.135cm} &\dotfill &  \dotfill &         - &  \dotfill &  \dotfill &        82 &  \dotfill &          \dotfill & ~\dotfill & ~\dotfill & ~\dotfill&  \dotfill&  \dotfill &                 \dotfill &  \dotfill\\
          68$\ast$ &\dotfill &  \dotfill &         - &  \dotfill &  \dotfill &         - &       177 &  16274709-2445350 &    ~15.75 &    ~12.81 &    ~11.13&     ~2.94&     ~1.68 &                    GY352 &        II\\
69\hspace{0.135cm} &\dotfill &  \dotfill &         - &  \dotfill &  \dotfill &        83 &  \dotfill &          \dotfill & ~\dotfill & ~\dotfill & ~\dotfill&  \dotfill&  \dotfill &                 \dotfill &  \dotfill\\
70\hspace{0.135cm} &\dotfill &  \dotfill &         - &  \dotfill &  \dotfill &        84 &  \dotfill &          \dotfill & ~\dotfill & ~\dotfill & ~\dotfill&  \dotfill&  \dotfill &                 \dotfill &  \dotfill\\
71\hspace{0.135cm} &\dotfill &       C11 &         - &  \dotfill &  \dotfill &  \dotfill &       182 &  16275180-2431455 &    ~14.68 &    ~11.19 &     ~8.71&     ~3.49&     ~2.48 &                    IRS54 &         I\\
72\hspace{0.135cm} &\dotfill &  \dotfill &         - &         - &       F32 &         - &       183 &  16275191-2446296 &    ~14.05 &    ~11.61 &    ~10.37&     ~2.44&     ~1.24 &                    GY377 &       III\\
73\hspace{0.135cm} &\dotfill &        54 &         - &  \dotfill &       F33 &        85 &       184 &  16275209-2440503 &    ~10.00 &     ~8.72 &     ~8.13&     ~1.28&     ~0.59 &                    IRS55 &       III\\
          74$\ast$ &\dotfill &  \dotfill &         - &         - &  \dotfill &         - &       186 &  16275565-2444509 &    ~12.34 &    ~11.15 &    ~10.47&     ~1.19&     ~0.68 &                    GY398 &      nIII\\
          75$\ast$ &\dotfill &  \dotfill &         - &  \dotfill &  \dotfill &         - &  \dotfill &          \dotfill & ~\dotfill & ~\dotfill & ~\dotfill&  \dotfill&  \dotfill &                 \dotfill &  \dotfill\\
76\hspace{0.135cm} &\dotfill &       C13 &         - &  \dotfill &  \dotfill &         - &       188 &  16275782-2440017 &    ~12.65 &    ~10.73 &     ~9.87&     ~1.92&     ~0.86 &                    GY410 &       III\\
77\hspace{0.135cm} &\dotfill &        55 &         - &         - &       F34 &         - &         - &  16275996-2448193 &    ~10.81 &     ~9.76 &     ~9.27&     ~1.05&     ~0.49 &        BKLT162800-244819 &      nIII\\
          78$\ast$ &\dotfill &         - &  \dotfill &  \dotfill &  \dotfill &         - &  \dotfill &  16280464-2434560 &    ~16.92 &    ~13.18 &    ~10.97&     ~3.74&     ~2.21 &                    GY463 &      nIII\\
          79$\ast$ &\dotfill &         - &  \dotfill &  \dotfill &  \dotfill &         - &  \dotfill &  16280478-2437100 &    ~13.62 &    ~11.45 &    ~10.34&     ~2.17&     ~1.11 &                    GY465 &      nIII\\
          80$\ast$ &\dotfill &         - &  \dotfill &         - &  \dotfill &         - &         - &          \dotfill & ~\dotfill & ~\dotfill & ~\dotfill&  \dotfill&  \dotfill &                 \dotfill &  \dotfill\\
          81$\ast$ &\dotfill &         - &  \dotfill &         - &  \dotfill &         - &         - &          \dotfill & ~\dotfill & ~\dotfill & ~\dotfill&  \dotfill&  \dotfill &                 \dotfill &  \dotfill\\
          82$\ast$ &\dotfill &         - &  \dotfill &         - &  \dotfill &         - &  \dotfill &  16280810-2445121 &    ~14.12 &    ~12.86 &    ~12.15&     ~1.26&     ~0.71 &                    GY478 &      nIII\\
          83$\ast$ &\dotfill &         - &  \dotfill &         - &  \dotfill &         - &  \dotfill &          \dotfill & ~\dotfill & ~\dotfill & ~\dotfill&  \dotfill&  \dotfill &                 \dotfill &  \dotfill\\
          84$\ast$ &\dotfill &         - &  \dotfill &         - &  \dotfill &         - &  \dotfill &          \dotfill & ~\dotfill & ~\dotfill & ~\dotfill&  \dotfill&  \dotfill &                 \dotfill &  \dotfill\\
          85$\ast$ &\dotfill &         - &  \dotfill &  \dotfill &  \dotfill &         - &  \dotfill &          \dotfill & ~\dotfill & ~\dotfill & ~\dotfill&  \dotfill&  \dotfill &                 \dotfill &  \dotfill\\
86\hspace{0.135cm} &\dotfill &         - &        33 &         - &  \dotfill &         - &  \dotfill &  16282373-2441412 &    ~11.73 &     ~9.83 &     ~9.01&     ~1.90&     ~0.82 &        BKLT162823-244140 &      nIII\\
87\hspace{0.135cm} &\dotfill &         - &        34 &         - &       F37 &         - &  \dotfill &  16282516-2445009 &     ~6.72 &     ~6.60 &     ~6.51&     ~0.12&     ~0.09 &        HD148352$\dagger$ &  \dotfill\\
\hline
\end{tabular}
\scriptsize

(1){\it XMM-Newton} source ID number, ``Rho Oph X-ray sources, Newton''. The asterisks indicate new X-ray sources found in this observation;
(2$-$9) The identification with source catalogs of (2) {\it Einstein} (Montmerle et al.\ 1983), ROX$-$; {\it $\dagger$ Grosso (2001) shows that one of these {\it Einstein} observation has detected an X-ray source associated with the YLW16 IR source group}; (3){\it ROSAT PSPC} (Casanova et al.\ 1995), ROXR1$-$; (4){\it ROSAT PSPC} (Casanova 1994), ROXR2$-$; (5){\it ASCA} (Katama et al.\ 1997), ROXA$-$; (6){\it ROSAT HRI} (Grosso et al.\ 2000), ROXR$-$; (7){\it Chandra} (Imanishi et al.\ 2001a); (8){\it ISO ISOCAM} (Bontemps et al.\ 2001), (9) 2MASS.
The dotted lines indicate that the {\it XMM-Newton} sources have no known counterpart.
The hyphens indicate that the {\it XMM-Newton} sources are outside the field of view of the instruments.
(10$-$14)If only the lower limits of the values are available, ``$>$'' is put before the values.
(15) For the source naming convention, see Andr\'e \& Montmerle (1994) and references therein.
The source names beginning with BKLT are from Barsony et al.\ (1997). 0600-2053 is from the catalog USNO-A1.0 (Monet et al.\ 1996). $\ddagger$ indicate foreground stars.
(16) The classifications of Class I, II, and III are from Bontemps et al.\ (2001) or Andr\'e \& Montmerle (1994).
The Class III candidates proposed in this paper are indicated by "nIII".
The bona fide brown dwarfs are also indicated in this column by "bd".
\label{irtable}
\end{table*}

\subsection{X-ray light curves}

We extract background-subtracted X-ray light curves for the ROXN sources.
X-ray counts are selected from circular regions around the sources where 90\% of the X-ray counts
 from the sources are included at 1.5 keV.\footnote{See calibration files XRT1\ XPSF\ 0005.CCF, XRT2\ XPSF\ 0005.CCF, and XRT3\ XPSF\ 0004.CCF.}
In crowded regions, we adjust the circle radius to avoid contamination by neighbouring sources.
Background X-ray counts for each source are extracted from annular regions with a 60$\arcsec$ inner radius and a 150$\arcsec$ outer radius around the sources.  
If the annuli contain other X-ray sources, regions within a 40$\arcsec$ radius 
around those sources  are excluded from the corresponding annuli.
To avoid decreasing too much the area of the annuli by excluding nearby X-ray sources, 
the outer radius was adjusted to keep a constant area for the background extraction region.

Since YSOs are known in general to exhibit X-ray flares, 
 we systematically look for flares in our sources.
We find that nearly half of the sources (40) have enough X-ray counts ($N > 150$, where $N$ is the sum of MOS1, MOS2 and PN counts)
to search for X-ray flares, using the following method.
We extract X-ray light curves with a uniform reference time binsize. A bin-to-bin variation was then defined to be a flare
if the X-ray counts in a given time bin are 1.5 times higher than at least one of the three preceding bins
 with a statistical significance above 4 $\sigma$.
We apply this method varying the reference time binsizes from 1000~s to 4000~s by steps of 1000~s, which covers the typical rise times of X-ray flares from YSOs.
To plot the light curves, we use an adaptive binning where the time binsize is tuned to keep a constant number of counts inside each time bin.
This enables us to catch any rapid time variability such as rising phase of X-ray flares. 
Depending on the total number of counts, the time bins are tuned so as
 to have enough counts in each bin to see the overall time profile properly (see tuning given in Table 3).

\begin{figure*}
\begin{center}
\resizebox{15cm}{!}{\includegraphics[angle=0]{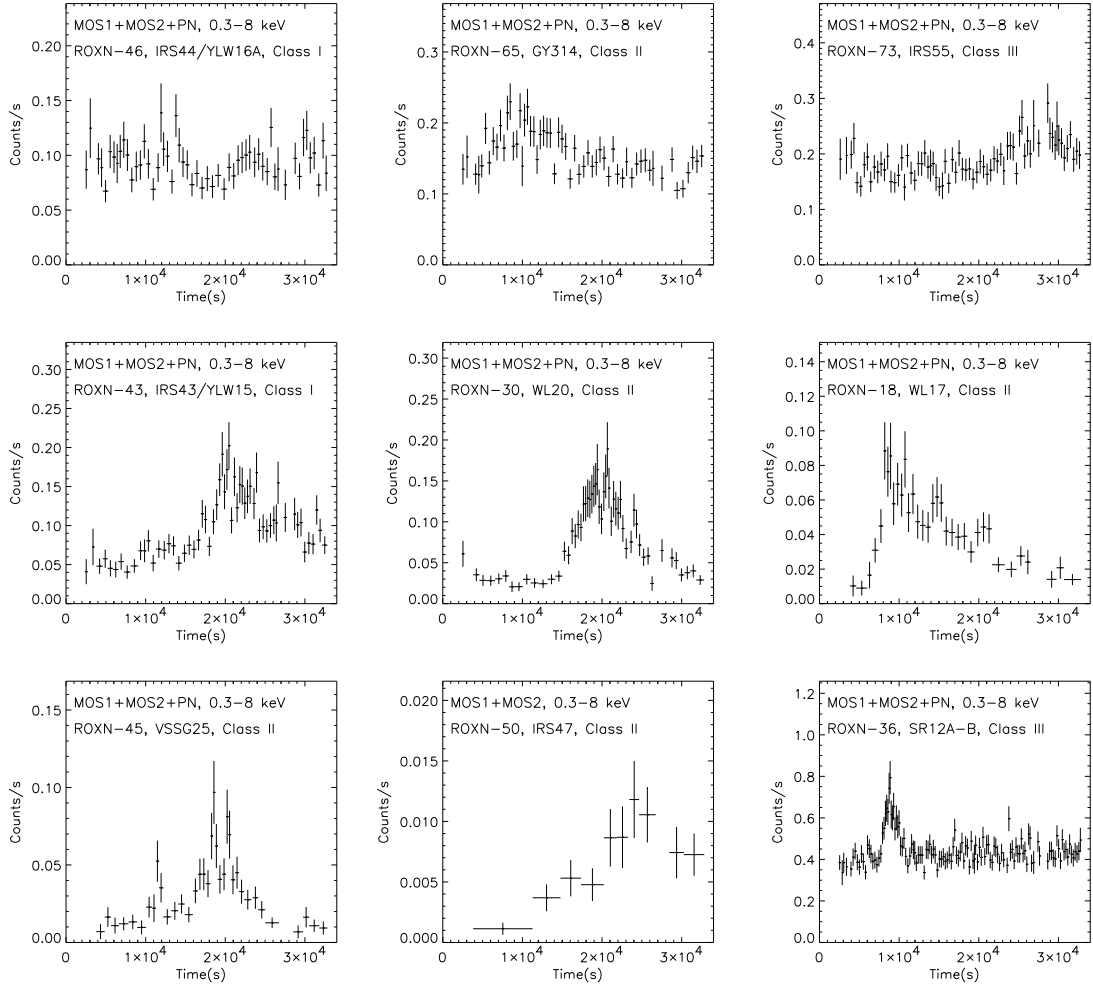}}
\caption{
A sample of X-ray background subtracted light curves of YSOs obtained with {\it XMM-Newton} showing variabilities.
High background time intervals were suppressed (holes in the light curve). 
We use an adaptive binning to keep a constant count number in each bin (see Table 3).
X-ray light curves of bona fide brown dwarfs, GY310 and GY141, are shown in Fig.~\ref{fig_lc_bd}.
}
\label{fig_lc}
\end{center}
\end{figure*}

Fig.~\ref{fig_lc} shows a sample of background-subtracted X-ray light curves.
The X-ray flares seen in Fig.~\ref{fig_lc} show a variety of profiles, broadly characterized by a fast increase and a roughly exponential decay, although some profiles are more symmetrical, with comparatively slow rise and decay phases of similar durations. Contrary to previous X-ray observations which had limited statistics, a ``typical" YSO X-ray flare profile cannot, in general, be reduced to a simple fast rise followed by an exponential decay. Such a complex behavior is also seen in the large X-ray flare database obtained in 40 ks {\it Chandra} observations of the Orion Nebula Cluster (Feigelson et al.\ 2002a).

In our $\rho$~Oph observations, X-ray flares are detected from 1/4 Class I, 6/13 Class II, and 3/12 Class III sources. 
These ratios are 2--3 times lower than those obtained by {\it Chandra} (Imanishi et al.\ 2001a; 14/18 for Class I + Class I candidates, 12/20 for Class II+III), which is to be expected since our {\it XMM-Newton} exposure (33 ks) is three times shorter than that of {\it Chandra} (100 ks).
The unclassified sources did not show any strong flares.

\begin{table}
\begin{center}
\caption{Tuning of the adaptive binning. Counts inside a time bin for a given source total counts.}

\begin{tabular}{ccc}
\hline
Source total counts  & Counts inside a time bin \\
\hline
$~~~0$ -- $~500$ & 15 \\
$~500$ -- $1000$ & 25 \\
$1000$ -- $2000$ & 35 \\
$2000$ -- $4000$ & 50 \\
$4000$ -- $6000$ & 80 \\
$6000  <  ~~~~~$      & 100 \\
\hline
\end{tabular}
\end{center}
\end{table}

\subsection{X-ray spectra}\label{subsec_spec}

\begin{figure*}
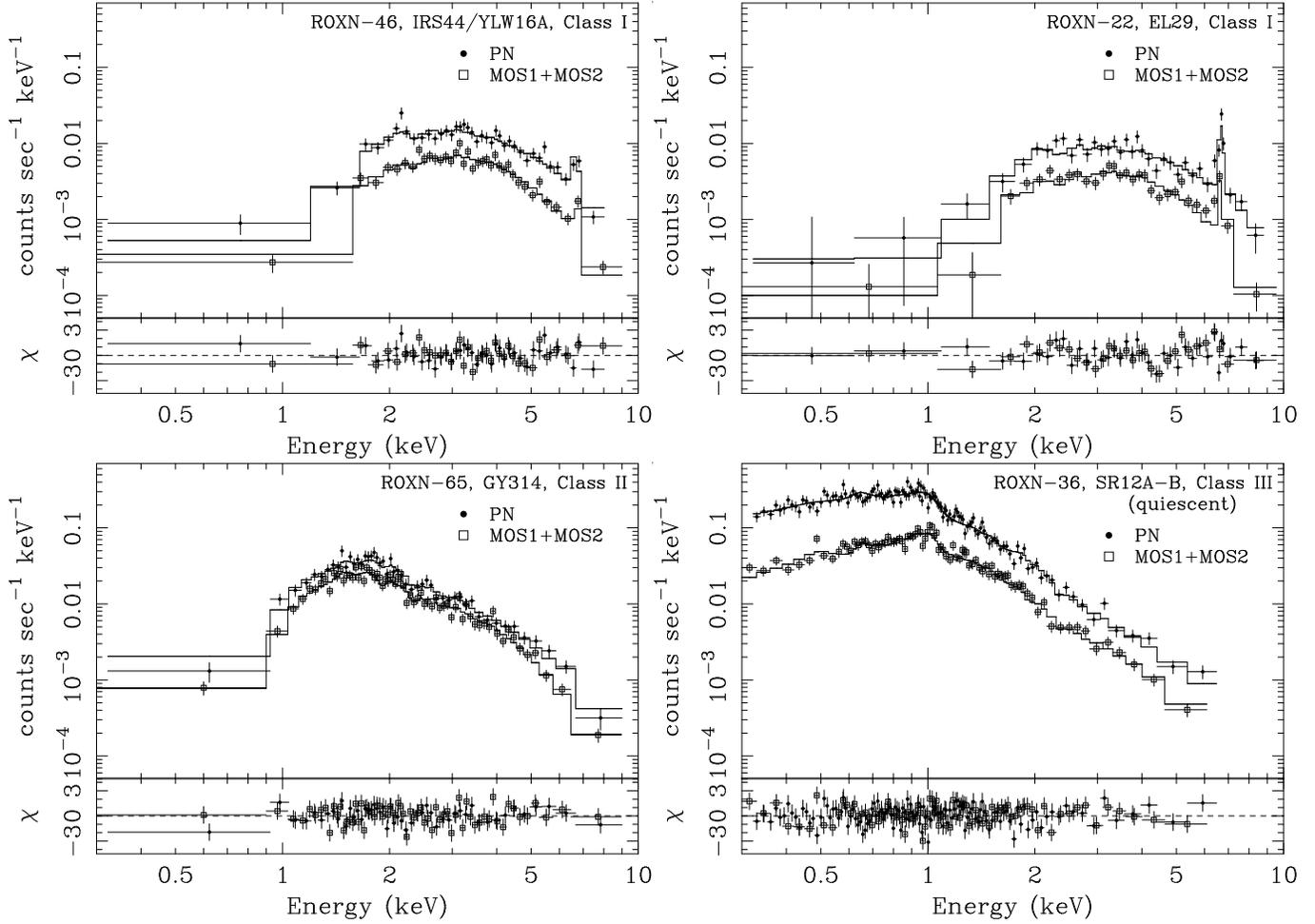

\resizebox{\hsize}{!}{\includegraphics[angle=270]{figure3a.ps},\includegraphics[angle=270]{figure3b.ps}}
\resizebox{\hsize}{!}{\includegraphics[angle=270]{figure3c.ps},\includegraphics[angle=270]{figure3d.ps}}
\caption{
X-ray spectra of IRS44/YLW16A (Class I), EL29 (Class I), GY314 (Class II), and SR12A$-$B (Class III) in the quiescent state obtained with {\it XMM-Newton}. In each panel, the filled circles and open squares indicate PN and MOS1$+$MOS2 spectra.
The solid lines show the best fit models whose spectral parameters are listed in Table 4.
}
\label{spec_fig}
\end{figure*}

We now compute background-subtracted X-ray spectra for sources with $N > 150$ cts (4 Class I, 13 Class II, 12 Class III sources, and 11 unclassified sources).
The way to extract X-ray counts and background counts is the same as for extracting the light curves.
\footnote{For ROXN-1, we exclude a circular region of 17.5$^{\prime\prime}$ radius centered on the source
 to keep the spectrum from being affected by pile-up.}
Since the difference between the responses of MOS1 and MOS2 is not large because they use the same type of CCD array,
 we add the MOS1 and MOS2 spectra to improve the statistics, i.e.,
we add the ancillary response files, and average the photon redistribution matrixes of MOS1 and MOS2 weighted by the exposure time of each detector. We treat the PN spectra separately.
Fig.~\ref{spec_fig} shows examples of the X-ray spectra obtained in our observation.

We first fit the separate PN and MOS1+MOS2 spectra with a single temperature thin thermal emission model (MEKAL) combined with an absorption model (WABS)\footnote{ We check consistency  of $N_\mathrm{H}$ determination between {\it Chandra} and {\it XMM-Newton} in Appendix B. We mention also uncertainties of X-ray derived $N_\mathrm{H}$ values especially for an effect of metal abundances for the absorption model.}. For sources which do not have enough counts to yield the plasma metal abundance,  we fix it to 0.3 solar,
a typical value for YSOs obtained in $\rho$ Oph with {\it ASCA} (Kamata et al.\ 1997) and {\it Chandra} (Imanishi et al.\ 2001a), and also in other star-forming regions. For calculations of emission measures and luminosities, we adopt $d = 140$ pc, based 
on {\it Hipparcos} measurements (see discussion in Bontemps et al.\ 2001), and for consistency with previous {\it ROSAT} study (Grosso et al.\ 2000).

As a result, we obtain acceptable one-temperature spectral fits for all the sources except for ROXs20A + ROXs20B, SR12A$-$B and IRS55 (see below).
The values of the fitting parameters are listed in Table 4.
The individual metal abundances obtained from the fits are found to be distributed around 0.3 solar, which is consistent with most existing YSO
 X-ray observations.
If we alternatively choose to fix the metal abundance to 0.3 solar for all the sources,
 we also obtain acceptable one-temperature fits with better constraints on other spectral parameters
  except for the Class I source EL29 (ROXN-22).
EL29 did not show X-ray flares during our observation, whereas it showed X-ray flares during the {\it ASCA} and {\it Chandra} observations (Kamata et al.\ 1997, Imanishi et al.\ 2001a).
We find 1.0 (0.80$-$1.3) for the metal abundance of EL29 relative to solar, 
which is far from the value, $\sim$0.3, observed for the quiescent level by {\it ASCA} and {\it Chandra} (Kamata et al.\ 1997, Imanishi et al.\ 2001a).
Actually, the K$\alpha$ line from He-like iron ions at 6.7 keV is
 clearly seen in the spectra of EL29 (see Fig.~\ref{spec_fig}). 
We note a residual in the spectral fitting around 6.4 keV in both PN and MOS1+MOS2, however
 adding a gaussian line at this energy does not change the metallicity at all.
This suggests that the metal abundances of the X-ray emitting plasma in YSOs could be variable.
For example, such metalicity enhancement could be explained by photospheric evaporation produced
 by flare (G\"udel et al.\ 2001).
Then we decide to list the values of the metal abundances obtained from X-ray spectral fitting 
 in our observation in Table 4.

Since it is often found that X-ray spectra with low statistics can be
   modeled by either a single temperature or a two-temperature MEKAL $+$ absorption model, 
 we also test a two-temperature MEKAL $+$ absorption model for all the ROXN sources. The metal abundance is then fixed to 0.3 solar except for EL29.
 We thus obtain acceptable two-temperature fits for all the sources, but the emission measures of the soft components are either extremely large or very low 
 for most sources. This means that the large absorption of the sources does not allow us to constrain the parameters of the soft component, so that one-temperature fits are sufficient in practice. 
However, for the bright Class III sources ROXs20A and ROXs20B (i.e., ROXN-27 and ROXN-28, which are spatially resolved but with too few counts, hence we use a circular area encompassing both sources to obtain a spectrum), SR12A$-$B (ROXN-36) and IRS55 (ROXN-73), two-, three-, and two-temperature MEKAL $+$ absorption models, respectively, are needed 
to obtain acceptable fits (see Fig.~\ref{spec_fig}).

\begin{table*}
\caption{Spectral parameters obtained from fit}
\tiny
\begin{tabular}{cccccccccc}
 \multicolumn{2}{l}{Class I} &&&&&&& \\
\hline 
ROXN  & NAME & & $kT$  & $N_{\rm{H}}$          & abundance & $E.M.$                 & $L_{\rm{X}}$             & $\chi^2$/d.o.f \\
      &&&       & ($10^{22}$ cm$^{-2}$) &           &($10^{52}$ cm$^{-3}$) & ($10^{29}$ erg s$^{-1}$) & \\ 
(1) & (2) & (3) & (4) & (5) & (6) & (7) & (8) & (9) \\
\hline \\
 22 & EL29   & & 4.3(3.6$-$5.1) & 4.9(4.4$-$5.3) & 1.0(0.80$-$1.3) & 16.0(13.6$-$18.9) & 27.5&  90.0/ 63.0 \\ 
 43 & IRS43/YLW15 & & 3.9(3.0$-$5.3) & 3.3(2.9$-$3.8) & [0.30] & 13.1(10.2$-$17.3) & 17.3&  36.5/ 35.0 \\ 
    &        &f& 3.2(2.7$-$3.9) & 4.0(3.5$-$4.5) & [0.30] & 29.1(22.8$-$37.0) & 34.8&  56.3/ 45.0 \\ 
 46 & IRS44/YLW16A  & & 2.7(2.4$-$3.1) & 5.3(4.9$-$5.7) & 0.28(0.19$-$0.38) & 45.6(38.3$-$53.9) & 49.7&  77.0/ 72.0 \\ 
 66 & IRS51  & & 2.1(1.8$-$2.8) & 4.2(3.5$-$4.8) & 0.62(0.27$-$1.18) & 13.8(10.2$-$18.5) & 16.2&  51.7/ 48.0 \\ 
\hline \\

\\
 \multicolumn{2}{l}{Class II}&&&&&&& \\
\hline 
ROXN &  NAME && $kT$  & $N_{\rm{H}}$          & abundance & $E.M.$                 & $L_{\rm{X}}$             & $\chi^2$/d.o.f \\
 ID    &&&       & ($10^{22}$ cm$^{-2}$) &           &($10^{52}$ cm$^{-3}$) & ($10^{29}$ erg s$^{-1}$) & \\ 
(1) & (2) & (3) & (4) & (5) & (6) & (7) & (8) & (9) \\
\hline \\
  1 & DoAr25& & 6.4(4.9$-$8.8) & 1.1(0.96$-$1.2) & 0.26(0.02$-$0.51) & 157(141$-$177) & 247 &  80.1/ 72.0 \\
 11 & SR24S & & 2.3(2.0$-$2.7) & 1.2(1.0$-$1.3)  & 0.24(0.07$-$0.43) & 16.3(13.4$-$19.4) & 15.9& 107.7/ 75.0 \\ 
 18 & WL17  & & 2.9(2.0$-$4.3) & 4.1(3.1$-$5.2)  & [0.30]         & 8.4(5.0$-$14.6) & ~9.6&  27.5/ 19.0 \\ 
    &       &f& 3.2(2.5$-$4.2) & 4.5(3.8$-$5.4)  & 0.66(0.31$-$1.2) & 18.1(13.1$-$25.7) & 24.8&  35.4/ 25.0 \\ 
 29 & IRS34 & & 1.8(1.1$-$3.7) & 2.7(1.6$-$4.4)  & [0.30]         & 12.8($<$41.5)  & 11.7 &   4.7/  5.0 \\ 
 30 & WL20  & & 2.3(1.8$-$3.0) & 2.0(1.6$-$2.4)  & [0.30]         & 8.1(5.9$-$11.4) & ~8.1&  16.2/ 17.0 \\ 
    &       &f& 3.4(2.7$-$4.5) & 2.0(1.7$-$2.3)  & 0.34(0.04$-$0.75) & 16.4(13.3$-$20.2) & 20.4&  34.7/ 29.0 \\ 
 33 & GY245 & & 2.1(1.3$-$5.0) & 7.7(5.2$-$11.0) & [0.30]         & 4.5(1.5$-$16.4) & 4.4&  10.9/  9.0 \\ 
 45 & VSSG25&f& 2.0(1.6$-$2.5) & 1.1(0.9$-$1.4)  & [0.30]         & 6.3(4.7$-$8.2) & ~5.9&  29.1/ 24.0 \\ 
 50 & IRS47 &f& 2.9(1.7$-$7.4) & 2.3(1.5$-$3.3)  & [0.30]         &  7.5($<$15.5)  & ~8.5 &   1.2/  2.0 \\ 
 57 & GY292 & & 2.3(2.1$-$2.5) & 1.7(1.5$-$1.8)  & 0.23(0.14$-$0.36) & 31.2(28.6$-$35.8) & 30.8& 159.3/153.0 \\ 
 61$^a$ & IRS49 & & 2.0(1.9$-$2.2) & 1.6(1.4$-$1.7)  & 0.18(0.08$-$0.29) & 27.1(23.9$-$30.7) & 24.1& 131.1/114.0 \\ 
 62 & GY310 &f& 2.5(1.7$-$3.6) & 0.82(0.59$-$1.2)  & [0.30]         & 1.8(1.4$-$2.4) & ~1.9&  22.4/ 13.0 \\ 
 63 & GY312 & & 2.3(1.7$-$3.3) & 7.6(6.0$-$8.7)  & [0.30]         & 11.4(6.4$-$22.5) & 11.6&  25.0/ 17.0 \\ 
 65 & GY314 & & 2.1(2.0$-$2.3) & 1.5(1.4$-$1.6)  & 0.16(0.07$-$0.26) & 31.6(28.4$-$35.0) & 28.4& 113.2/113.0 \\ 
\hline \\

\\
 \multicolumn{2}{l}{Class III}&&&&&&& \\
\hline 
ROXN  &  NAME && $kT$  & $N_{\rm{H}}$          & abundance & $E.M.$                 & $L_{\rm{X}}$             & $\chi^2$/d.o.f \\
       &&&       & ($10^{22}$ cm$^{-2}$) &           &($10^{52}$ cm$^{-3}$) & ($10^{29}$ erg s$^{-1}$) & \\ 
(1) & (2) & (3) & (4) & (5) & (6) & (7) & (8) & (9) \\
\hline \\
5  &  GY112      & & 0.98(0.79$-$1.1) & 0.36(0.31$-$0.50) & 0.10(0.06$-$0.16) & 7.5(6.1$-$11.8) & 4.3&  40.9/ 37.0 \\ 
14$^a$ &  GY193  & & 1.27(1.19$-$1.74) & 0.87(0.56$-$1.07) & [0.30] & 7.0(4.6$-$8.2) & 6.0&  29.4/ 38.0 \\ 
15 &  GY194      & & 1.5(1.2$-$1.9) & 0.79(0.6$-$1.1) & [0.30] & 3.2(2.6$-$3.9) & 2.8&  20.7/ 13.0 \\ 
25 &  WL19       & & 2.5(1.5$-$4.6) & 8.62(6.0$-$12.8)  & [0.30] & 9.2(3.7$-$30.7) & 9.8&   8.4/ 10.0 \\ 
27,28$^{a,b}$ & ROXs20A, ROXs20B & & & 0.13(0.06$-$0.18) & 0.42(0.23$-$1.21) &         & 4.6&  59.2/ 40.0 \\ 
      &          & & 0.66(0.56$-$0.77) &                &           & 1.7(0.8$-$2.2) & 1.9& \\ 
      &          & & 1.66(1.30$-$2.29) &                &           & 2.8(1.9$-$3.9) & 2.8& \\ 
32,34,35$^{b,c}$ & WL5,WL4,WL3 &           &2.5(2.2$-$3.0) & 4.7(4.2$-$5.2) & [0.30] & 60.4(47.4$-$76.9) & 64.2& 104.6/ 74.0 \\ 
36 & SR12A$-$B & &                   & 0.11(0.08$-$0.13) & 0.12(0.08$-$0.17) &                   & 21.9& 248.6/205.0\\ 
   &           & & 0.25(0.23$-$0.27) &                   &                   & 18.2(11.1$-$30.2) & 4.4& \\ 
   &           & & 0.97(0.89$-$1.03) &                   &                   & 18.6(17.0$-$25.2) & 11.1& \\ 
   &           & & 3.5(2.5$-$6.5)    &                   &                   & 5.6(3.1$-$8.3) & 6.5& \\ 
   &           &f&                   & 0.06(0.03$-$0.10) & 0.24(0.13$-$0.55) &                   & 24.0& 74.0/ 60.0 \\ 
   &           & & 0.32(0.28$-$0.38) &                   &                   & 7.8(4.3$-$17.1) & 4.2& \\ 
   &           & & 1.22(1.09$-$1.35) &                   &                   & 15.3(5.5$-$22.5) & 12.0& \\ 
   &           & & 12.0($>$3.1)      &                   &                   & 4.9(3.0$-$10.4) & 7.7& \\ 
39 & GY253     & & 1.6(1.4$-$2.0)    & 4.5(3.9$-$5.2)   & [0.30]         & 24.9(17.5$-$34.9) & 22.0&  57.7/ 52.0 \\ 
52 & BKLT162730-244726 &  & 1.9(1.3$-$2.3) & 1.0(0.8$-$1.5) & [0.30] & 6.6(5.3$-$11.2) & 6.1&  44.1/ 43.0 \\ 
72 & GY377     & & 1.8(1.1$-$2.7)    & 1.1(0.7$-$1.8)    & [0.30]         & 1.7(0.9$-$3.6) & 1.5&   7.1/ 11.0 \\ 
73 & IRS55     & &                   & 1.1(1.0$-$1.2)   & 0.31(0.20$-$0.50) &                   & 62.3& 139.3/141.0 \\ 
   &           & & 0.64(0.60$-$0.68) &                   &                  & 46.9(28.0$-$67.2) & 45.7& \\ 
   &           & & 2.5(2.2$-$3.0)    &                   &                  & 15.7(12.7$-$19.2) & 16.6& \\ 
76 &  GY410    & & 1.8(1.3$-$2.4)    & 1.5(1.1$-$1.9)   & [0.30]         & 3.8(2.5$-$5.9) & 3.4&   8.9/ 23.0 \\ 
\hline \\

\\
 \multicolumn{2}{l}{New Class III candidates} &&&&&&& \\
\hline 
ROXN  & NAME && $kT$  & $N_{\rm{H}}$          & abundance & $E.M.$             & $L_{\rm{X}}$         & $\chi^2$/d.o.f \\
 ID    &&&       & ($10^{22}$ cm$^{-2}$) &           &($10^{52}$ cm$^{-3}$) & ($10^{29}$ erg s$^{-1}$) & \\ 
(1) & (2) & (3) & (4) & (5) & (6) & (7) & (8) & (9) \\
\hline
31 & 16271643-24311 && 4.2(2.2$-$15.1)  & 5.2(3.5$-$7.4)  & [0.30] & 6.1(3.2$-$14.3) & 8.3&  25.8/ 26.0 \\ 
41 & GY259          && 3.4($>$1.3) & 2.2(1.1$-$4.5)  & [0.30] & 1.9(0.8$-$6.3) & 2.3&   0.5/  2.0 \\ 
44 & GY266          && 2.3(1.1$-$6.1)   & 2.8(1.6$-$5.0)  & [0.30] & 2.4(1.0$-$8.2) & 2.5&   7.5/  6.0 \\ 
53 & GY283          && 1.4(0.5$-$2.6)   & 0.47(0.2$-$1.5) & [0.30] & 0.5(0.3$-$3.0) & 0.5&   2.1/  4.0 \\ 
58 & GY295          && 1.9(1.2$-$2.8)   & 0.34(0.2$-$0.6) & [0.30] & 0.7(0.5$-$0.8) & 0.6&  12.3/ 11.0 \\ 
77 & BKLT162800-244819 && 0.5(0.4$-$0.7)   & 0.97(0.8$-$1.1) & [0.30] & 8.7(4.4$-$17.2) & 7.8&  24.2/ 25.0 \\ 
78 & GY463          && 1.6(1.2$-$2.1)   & 2.6(2.0$-$3.4)  & [0.30] & 11.0(6.8$-$20.2) & 9.6&  16.0/ 20.0 \\ 
86 & BKLT162823-244140 && 4.3(2.2$-$13.2)  & 0.87(0.5$-$1.3) & [0.30] & 4.0(2.8$-$6.7)  & 5.5 &   9.7/  9.0 \\ 

\hline \\

\\
 \multicolumn{4}{l}{X-ray sources with no optical or IR counterpart} &&&&& \\
\hline 
ROXN  & NAME && $kT$  & $N_{\rm{H}}$          & abundance & $E.M.$             & $L_{\rm{X}}$         & $\chi^2$/d.o.f \\
 ID    &&&       & ($10^{22}$ cm$^{-2}$) &           &($10^{52}$ cm$^{-3}$) & ($10^{29}$ erg s$^{-1}$) & \\ 
(1) & (2) & (3) & (4) & (5) & (6) & (7) & (8) & (9) \\
\hline
51 & \dotfill       && 10.5($>$3.9)& 0.92(0.6$-$1.4) & [0.30] & 2.5(2.0$-$3.8) & 4.4&  40.1/ 28.0 \\ 
64 & \dotfill       && 82.4($>$20.5) & 2.2(1.7$-$2.9)& [0.30] & 3.5(2.9$-$4.1) & 5.5&  64.5/ 43.0 \\ 
67 & \dotfill       && 3.6($>$1.4) & 9.5(4.8$-$18.0) & [0.30] & 6.7(2.4$-$46.9) & 8.6&  25.4/ 29.0 \\ 
\hline \\
\hline \\

\end{tabular}
\\
(1){\it XMM-Newton} source ID number, ``Rho Oph X-ray sources, Newton''.
(2) The source name from column (15) in Table 2.
(3) "f" means that the X-ray spectra are extracted during the flare phase.
(4$-$7) The 90\% confidence regions are given between parenthesis. 
(5) We use WABS for absorption model in XSPEC where standard solar metal abundances are used. The use of reviese solar abundances (Holweger 2001, Allend Prieto et al. 2001, 2001) would increase N$_\mathrm{H}$ by 20 $\sim$\% (Vuong et al. 2003).
(7),(8) {The emission measures and X-ray luminosities were calculated from the best-fit PN parameters assuming a distance of 140 pc. Those from MOS1+MOS2 parameters are consistent within errors.}\\
$^a$ These sources showed small flare during the observation, but the X-ray counts in the flares are not enough to make spectra.\\
$^b$ It is difficult to obtain individual spectrum without contamination from neighbouring sources.
Then we extracted X-ray spectrum from a circular region which contain all the sources.\\
$^c$ Since the X-ray counts from the source 32 is dominant, we put the spectral parameters in the table for Class III.

\label{spec_tab}
\end{table*}

\section{Comparison with {\it Chandra}  source detection}\label{sec_chan}

\begin{figure}
\resizebox{\hsize}{!}{\includegraphics[angle=270]{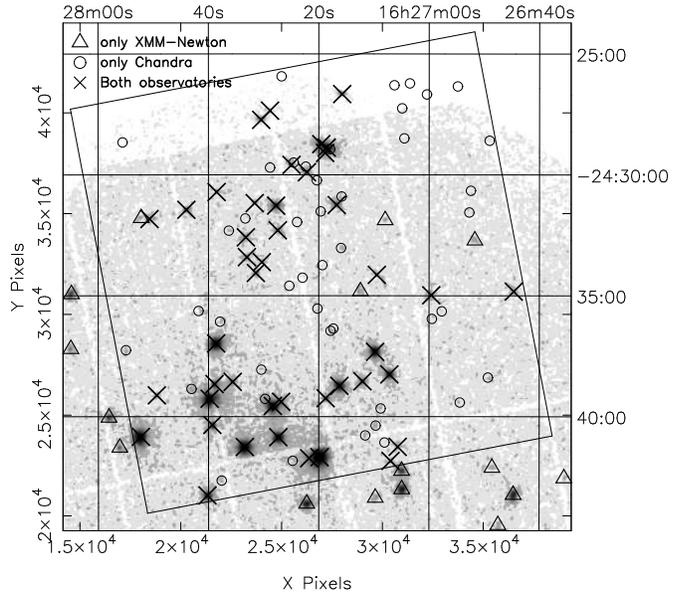}}
\caption{
Comparison of source detection between {\it XMM-Newton} and {\it Chandra} (Imanishi et al. 2001a).
The background image shows the {\it XMM-Newton} image in the 0.3$-$8 keV energy range smoothed with
 a 6\arcsec-FWHM Gaussian. 
The rectangle shows the field of view of {\it Chandra}.
The triangles, circles, and crosses show the positions of the X-ray sources detected with only {\it XMM-Newton}, only {\it Chandra}, and both observatories, respectively. 
}
\label{fig_im}
\end{figure}

{\it Chandra} observed the $\rho$ Ophiuchi cloud core F region with an exposure of 100 ks (Imanishi et al.\ 2001a), although, contrary to our {\it XMM-Newton} observations, the pointing was somewhat off-centered with respect to the peak of the CO emission of core F.
Fig.~\ref{fig_im} shows the {\it XMM-Newton} and {\it Chandra} X-ray source positions superimposed on the 0.3$-$8 keV band image obtained by {\it XMM-Newton}. 
{\it Chandra} detected in 100 ks 87 X-ray sources in its field of view (17$\arcmin$x17$\arcmin$). This the same number of sources as in our three times shorter {\it XMM-Newton} observations, in a field of view which is nearly four times as large. We shall return to this coincidence below.
In the region covered by the field of view of both observatories, 47 X-ray sources are detected with {\it XMM-Newton} and 81 X-ray sources are detected with {\it Chandra}, while 43 X-ray sources are detected  with both observatories. 

\begin{figure}
\resizebox{\hsize}{!}{\includegraphics[angle=90]{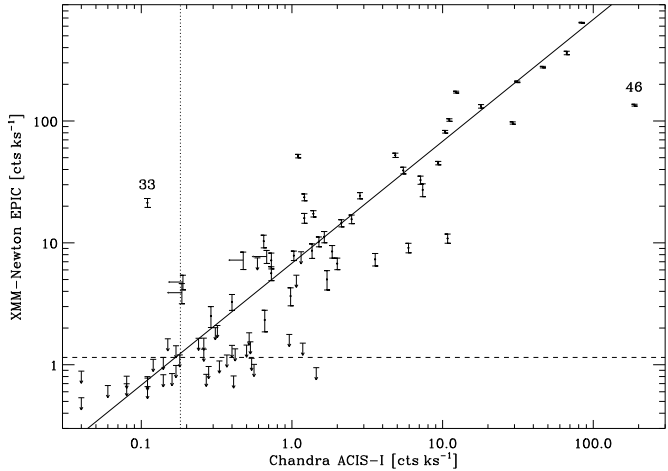}}
\caption{
Comparison between {\it XMM-Newton/}EPIC and {\it Chandra/}ACIS-I count rates for X-ray sources
 in {\it Chandra} and {\it XMM-Newton} overlapping field-of-views.
The continuous line shows the median of the count rate ratio for sources detected by both instruments.
The dashed line is the median of the count rate upper limit for {\it Chandra} sources which are not
 detected during the 33 ks {\it XMM-Newton} observation. The dotted line works the {\it Chandra} detection
 limit ($\sim$ 6 counts) for a 33 ks observation.
}
\label{fig_chanc}
\end{figure}

Fig.~\ref{fig_chanc} shows {\it XMM-Newton} vs. {\it Chandra} count rates for all these X-ray sources.
Upper limits correspond to the 99.9989\% confidence level threshold
 (i.e., 4.4 $\sigma$ level for Gaussian statistics) 
for the {\it Chandra} sources undetected by {\it XMM-Newton}, and for the {\it XMM-Newton} sources 
undetected by {\it Chandra}, using a method explained in Appendix A. 
The median of the count rate upper limits of sources detected with only {\it Chandra} 
 is the dotted horizontal line in Fig.~\ref{fig_chanc}, which indicates  
 also the 4.4 $\sigma$ detection threshold for {\it XMM-Newton}.
There is a good correlation between {\it XMM-Newton} and {\it Chandra} count rates.
The median of the ratios between {\it XMM-Newton} and {\it Chandra} count rates for X-ray sources detected by both observatories is 6.8, which is indicated by a continuous line in Fig.~\ref{fig_chanc}: ``the median ratio line".
Most of the data points are scattered close to this median ratio line. 
A few sources display large discrepancies between {\it XMM-Newton} and {\it Chandra} observations,
  which suggests that their X-ray luminosities could have changed due to time 
 variability.
We check the two most extreme cases, sources ROXN-33 (GY245) and ROXN-46 (YLW16A), which are far from the median ratio line.
GY245 did not show any flare during the {\it Chandra} observation, but showed a flare during the {\it XMM-Newton} observation, while YLW16A showed an X-ray flare during the {\it Chandra} observation, but did not show any flare during the {\it XMM-Newton} observation.
Hence, flaring activity explains well the large apparent discrepancies observed for these two sources.

There is a large difference between the number of X-ray sources detected with {\it XMM-Newton} and detected with {\it Chandra} in the region common to the field of view of both observatories, which we now seek to explain.
The observation time of {\it Chandra}, 100 ks, 
 is three times longer than that of {\it XMM-Newton}, 33 ks. 
If the observation time of {\it Chandra} had been as short as that of {\it XMM-Newton}, 
 the number of sources detected by {\it Chandra} would have been smaller. More precisely,
if we assume that $\sim$ 6 counts are needed to make a {\it Chandra} detection 
and that the count rates of the X-ray sources are constant,
 in a 33 ks {\it Chandra} observation, 
 only those sources with count rates above 0.18 counts ks$^{-1}$ would have been detected (see the dotted line in Fig.~\ref{fig_chanc}).
Fig.~\ref{fig_chanc} shows that most of the 23 X-ray sources detected by {\it Chandra} only
 with count rates above 0.18 counts ks$^{-1}$ are distributed well below the median ratio line,
  which indicates that these sources probably showed X-ray flares only during the 100 ks {\it Chandra} observation, 
 but could disappear in a shorter, 33 ks observation.
All in all, when the exposure time is set to about 30 ks for both observatories, and taking into account the differences in fields of view and detection thresholds, {\it Chandra} and {\it XMM-Newton} detect roughly the same
 number of sources.

In summary, we find that the respective merits of {\it XMM-Newton} and {\it Chandra}, for the same type of observations (here a typical low-mass star-forming region) are as follows. {\it XMM-Newton} has an effective area about three times larger than {\it Chandra}, 
so that {\it XMM-Newton} detects more source counts than {\it Chandra} for the same observation time:
 as shown in Fig.~\ref{fig_chanc}, the {\it XMM-Newton} count rate is 6.8 times larger than 
 the {\it Chandra} count rate.
On the other hand, {\it Chandra} has a lower background, and a sharper point spread function than {\it XMM-Newton}, 
 which means {\it Chandra} suffers less noise than {\it XMM-Newton} for point source detection. For relatively short exposures (but typical in a proposal request) of 30 ks, though these values depend on position with respect to the pointing axis,
  the signal-to-noise ratios are similar for both observatories.
For longer exposure times, {\it Chandra} provides a better detection ability than {\it XMM-Newton}, while for shorter exposure times, {\it XMM-Newton} provides a slightly better detection ability than {\it Chandra},
 because the statistics with {\it Chandra} is photon-dominated, while for {\it XMM-Newton} it is background-dominated. 
 Because it detects more photons in a given exposure, {\it XMM-Newton} allows a more detailed spectral and time variability analysis for moderately bright sources (above $\sim 150$ counts), typical of YSOs, than does {\it Chandra}.

\section{IR properties of the {\it XMM-Newton} sources}\label{sec_ir}

ISOCAM provides a reliable classification of YSOs in the $\rho$ Ophiuchi cloud in terms of Class I or II sources, from the detection of IR excess produced by circumstellar material (Bontemps et al.\ 2001). 
We detect X-rays from 7/11 Class I sources, 25/61 Class II sources, and 14/15 Class III sources in the  overlapping region between the field of view of {\it XMM-Newton} and the ISOCAM survey.
\footnote{Note that the prototype of Class 0 sources, VLA1623 (Andr\'e et al.\ 1993) is located in the dense core A, which is not covered by our observation.}
We find a high detection rate for Class I sources (64 \%), which is higher than that of Class II sources (48 \%).
The {\it XMM-Newton} detection rate of Class I sources is comparable to that obtained from the {\it Chandra} observation, 
but the detection rate for  Class II sources is lower by more than 70 \% than that obtained by {\it Chandra} (Imanishi et al.\ 2001a).
The Class III sample is biased because these sources were mostly identified by X-rays, which explains why their high detection rate (93 \%).

\begin{figure}
\resizebox{\hsize}{!}{\includegraphics[angle=0]{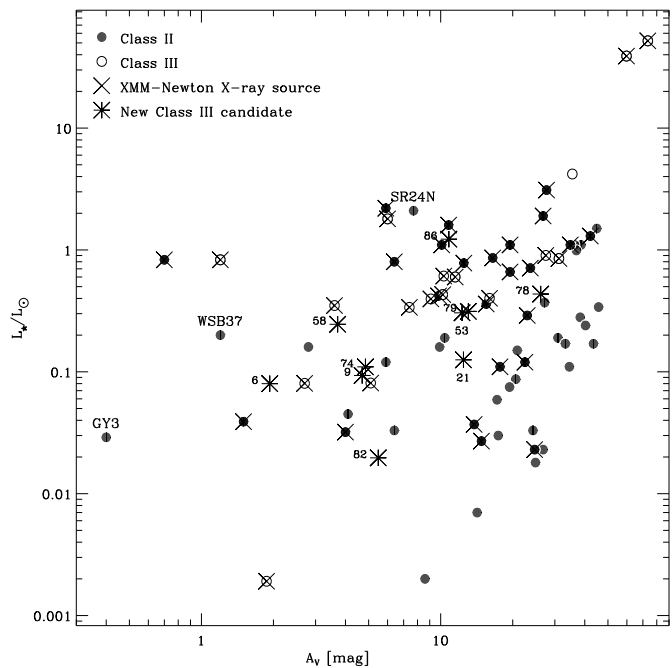}}
\caption{
Stellar luminosity vs. extinction for Class II (filled circles) and III (open circles)
 sources in the ISOCAM survey (Bontemps et al. 2001) and {\it XMM-Newton} overlapping area.
The sources which have {\it XMM-Newton} counterparts are indicated by crosses.
The new Class III candidates are indicated by asterisks and are labeled with ROXN numbers of Table 1 and 2.
}
\label{fig_av}
\end{figure}

Fig.~\ref{fig_av} is a scatter plot of stellar luminosities ($L_\star$) vs. extinction (A$_\mathrm{V}$), for Class II (filled circles) and Class III sources (open circles) from Bontemps et al.\ (2001) results. Sources having {\it XMM-Newton} X-ray counterparts are indicated by crosses.
 Most of the X-ray non-detected Class II sources are distributed in the lower-right region (low $L_\star$ and high A$_\mathrm{V}$) of Fig.~\ref{fig_av}.
Assuming that there is an $L_X/L_\star$ correlation (see below, Sect. 6),
 combined with the high extinction, the X-ray non-detected Class II sources in this region could simply lie below the detection threshold of {\it XMM-Newton} (see Grosso et al.\ 2000) which, as discussed above, is higher than for {\it Chandra}.

However, some X-ray non-detected Class II sources, like SR24N , WSB37, and GY3, 
 are not located in the lower-right region of Fig.~\ref{fig_av}. This could be explained in two ways.
First, their individual $L_X/L_\star$ ratios may be lower than typical ones.
Second, their extinction may be higher than given by the statistical extinctions estimated from the NIR data (Bontemps et al.\ 2001).
This could happen if the circumstellar disk of 
 these Class II sources is seen close to edge-on:
in such a case, the central star is not visible, and
 we just observe its scattered light so that a statistical value of A$_\mathrm{V}$ underestimates the real value.
Conversely, two luminous Class III sources with very high extinction, WL5 (ROXN-32) and WL19 (ROXN-25), are located in the upper-right corner of Fig.~\ref{fig_av}.
Their X-ray spectral fits provide a column density consistent with high absorption (see Table 4).
These two Class III sources are probably located {\it behind} the dense core F of  cloud. A similar conclusion, in the much wider field of view of {\it ROSAT}, was reached for many Class III sources by Grosso et al.\ (2000).
The correlation between X-ray absorption column density and optical extinction was studied in detail
 by Vuong et al.\ (2003), who compare the gas and dust properties of the dense interstellar matter in
 nearby star-forming regions using X-rays from Class III sources.

Considering that ISOCAM observations of Bontemps et al.\ (2001) were sensitive enough to detect IR excess, if any, from the new X-ray sources found in our observation, they could be Class III candidates.
There are 37 {\it XMM-Newton} X-ray sources without IR classification in Table 2,
 18 of which have 2MASS counterparts. With the exception of ROXN-10 (identified with the bona fide brown dwarf GY 141, see below, \S 5), 
we check now whether the near-IR photometry of these X-ray sources is consistent with that of Class III sources.

\begin{figure}
\resizebox{\hsize}{!}{\includegraphics[angle=0]{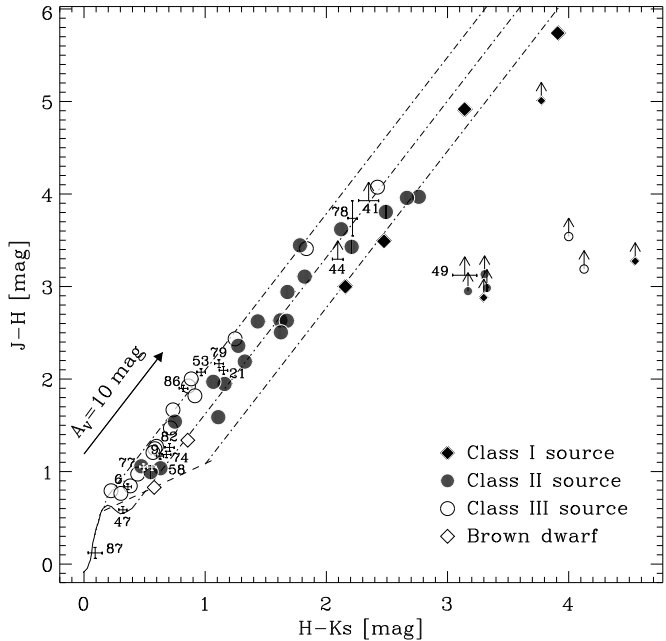}}
\caption{
Color-color diagram of the {\it XMM-Newton} sources which have 2MASS near-IR counterparts.
The arrow shows an extinction of 10 mag (Cohen et al. 1981).
The intrinsic colors of giants (dotted line) and A0$-$M6 dwarfs from Bessel \& Brett (1988),
adapted for the 2MASS photometric system using the 2MASS color transformation (Carpenter 2001; Cutri et al. 2003),
are plotted for comparison. The dashed line shows the locus of the intrinsic color classical T Tauri stars (Meyer et al. 1997).
Arrows indicate lower limits on $J-H$ when $J$ magnitude is not available.
}
\label{fig_cc}
\end{figure}

Fig.~\ref{fig_cc} shows the color-color diagram 
 of the {\it XMM-Newton} X-ray sources identified with the 2MASS near-IR sources.
Class I, II, and III sources and brown dwarfs are indicated by filled diamonds, filled circles, open circles, and open diamonds, respectively; and sources without IR classification are indicated
 by asterisks labeled with the ROXN numbers of Table 2.
As ROXN-31 is too faint to have measurable $J$ and $H$ magnitudes, it is removed from our sample, and
 16 unclassified sources are plotted in Fig.~\ref{fig_cc}.
For ROXN-41, -44, and -49,  only lower limits are available for the $J-H$ color.
ROXN-87, corresponding to {\it ROSAT} source ROXR-F37, was identified as a foreground F2 V star, HD148352 (Grosso et al.\ 2000), which is consistent with its position on the locus of the intrinsic colors (Bessel \& Brett 1988). This is also consistent with the fact that its X-ray hardness ratio is -1.0 for MOS1 and -1.0 for MOS2 from Table 1, which means there is no hard X-ray emission above 2 keV.
ROXN-47 is also located at the position of a late-type M dwarf without extinction in Fig.~\ref{fig_cc}; its hardness ratio is relatively low, $-$0.54, $-$1.00, and $-$0.76 for MOS1, MOS2, and PN, respectively. We thus consider this source also as a foreground star

\begin{figure*}
\resizebox{\hsize}{!}{\includegraphics[angle=0]{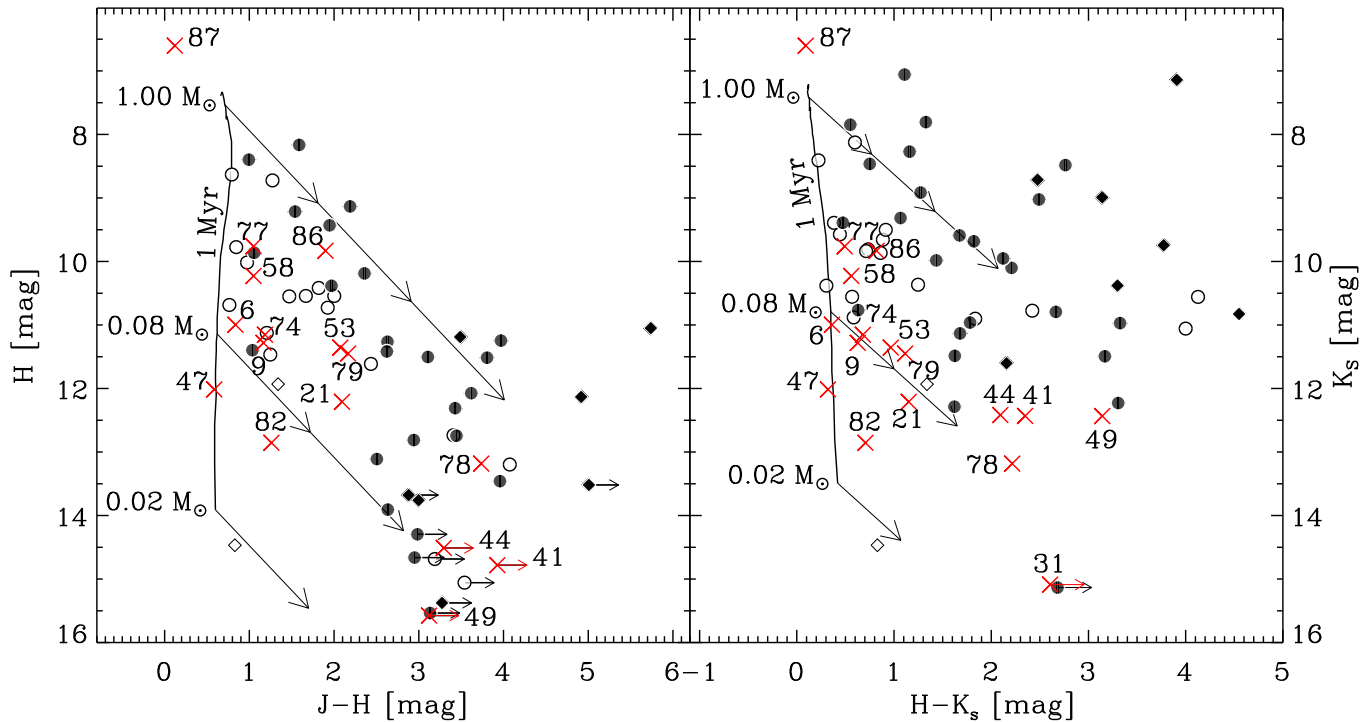}}
\caption{
Color-magnitude diagram of the {\it XMM-Newton} X-ray sources with 2MASS counterparts.
The solid line shows the 1 Myr isochrone for solar metalicity 
low mass stars from 0.02 to 1.00 M$_{\sun}$ (Baraffe et al. 1998), and 10 mag reddening vectors (Cohen et al. 1981).
Symbols are the same as in Fig.~\ref{fig_cc}. 
Arrows indicate lower limits on $J-H$ and $H-K_\mathrm{s}$ when $J$ and $H$ magnitudes are not available, respectively.
}
\label{fig_cm}
\end{figure*}

 Fig.~\ref{fig_cm} shows  the color-magnitude diagrams of $H$ vs. $J-H$ in the left panel, and $K_\mathrm{s}$ vs. $H-K_\mathrm{s}$
  in the right panel for {\it XMM-Newton} sources having 2MASS counterparts.
(Note that ROXN-31 appears only in the $K_\mathrm{s}$ vs. $H-K_\mathrm{s}$ diagram since this source is detected only in the $K_\mathrm{s}$ band.)
We plot for comparison the 1 Myr isochrone (Baraffe et al.\ 1998) and reddening vectors (Cohen et al.\ 1981).
The positions of the remaining 15 sources without IR classification are well mixed with those of well-known
 Class II and III sources.
Therefore, we propose these 15 X-ray sources as new Class III candidates (labeled `nIII' in Table 2), therefore doubling the present number of Class III sources in this area. 
Spectroscopic follow-up is now needed to determine the effective temperature of these objects, and to put them
 in an H-R diagram to confirm their pre-main sequence status. 
We estimate stellar luminosities and extinctions of these new Class III candidates using $J$ and $H$ band 2MASS magnitudes (see formulas in Bontemps et al.\ 2001), and plot them in Fig.~\ref{fig_av}.
The new Class III candidates are also mixed with the Class II and III sources in Fig.~\ref{fig_av}.

In summary, we find a high {\it XMM-Newton} detection rate for Class I and Class III sources, consistent with the {\it Chandra} results, and a lower detection rate for Class II sources. For those sources, the difference is probably due to a combination of high extinction (interstellar + circumstellar) and of lower detection efficiency of {\it XMM-Newton} compared with {\it Chandra}. We however detect 15 previously unknown X-ray sources, which we propose as new Class III candidates, pending spectroscopic follow-ups to confirm their nature. Our {\it XMM-Newton} observations thus allow for a significant improvement of the YSO census in the $\rho$~Oph cloud core F region (15 new YSOs in addition to a total of 87 previously known from X-ray/IR observations, and a potential doubling of the number of Class III sources).

\section{X-ray detection of young bona fide brown dwarfs}\label{sec_bd}  

Since {\it ROSAT} observations, it is known that young brown dwarfs also emit X-rays (Neuh{\"a}user \& Comer{\'o}n 1998; Neuh{\"a}user et al.\ 1999; Imanishi et al.\ 2001a; Imanishi et al.\ 2001b; Preibisch \& Zinnecker 2001; Mokler \& Stelzer 2002; Feigelson et al.\ 2002b; Tsuboi el al.\ 2003). We thus look for X-rays from young {\it bona fide} brown dwarfs in our observation, i.e., from young objects with substellar status confirmed by spectroscopy. There are in our field-of-view only three {\it bona fide} brown dwarfs\footnote{We note in Fig.~\ref{fig_cm} that ROXN-82 is located well below the substellar limit of $\sim$ 0.08\,M$_{\sun}$ and could be a brown dwarf candidate, but it was found to be a very low-mass M star by a near-IR spectroscopical follow-up (Grosso et al.\ 2004, in preparation).}: GY202, GY141, and GY310 (Mart{\'\i}n et al.\ 1999; Cushing et al.\ 2000; Wilking et al.\ 1999). Bontemps et al.\ (2001) classified GY310 as Class II, and recently Mohanty et al.\ (2004) detected it from the ground with {\it Subaru} at 8.6 and 11.7$\mu$m, confirming the presence of significant mid-infrared excess arising from an optically thick, flared dusty disk.

A low S/N ratio X-ray detection of GY202 was reported by Neuh{\"a}user et al.\ (1999) from the {\it ROSAT/PSPC} pointing observation of Casanova et al.\ (1995). We find that the identification of this {\it ROSAT/PSPC} source with GY202, located 19\arcsec~away, is dubious because the closest counterpart is in fact WL1, an embedded (A$_\mathrm{V} \sim 21$\,mag) Class II source (Bontemps et al.\ 2001), located only 9\arcsec~away. Moreover, in spite of the fact that we do not detect WL1, its X-ray emission is confirmed by the {\it Chandra} observation (source 13 of Imanishi et al.\ 2001a) with a luminosity roughly consistent with the {\it ROSAT/PSPC} estimate, whereas GY202 is not detected either by {\it Chandra} (Imanishi et al.\ 2001b; Imanishi et al.\ 2003) and {\it XMM-Newton} (this work).

\begin{figure}
\begin{center}
\resizebox{\hsize}{!}{\includegraphics[angle=0]{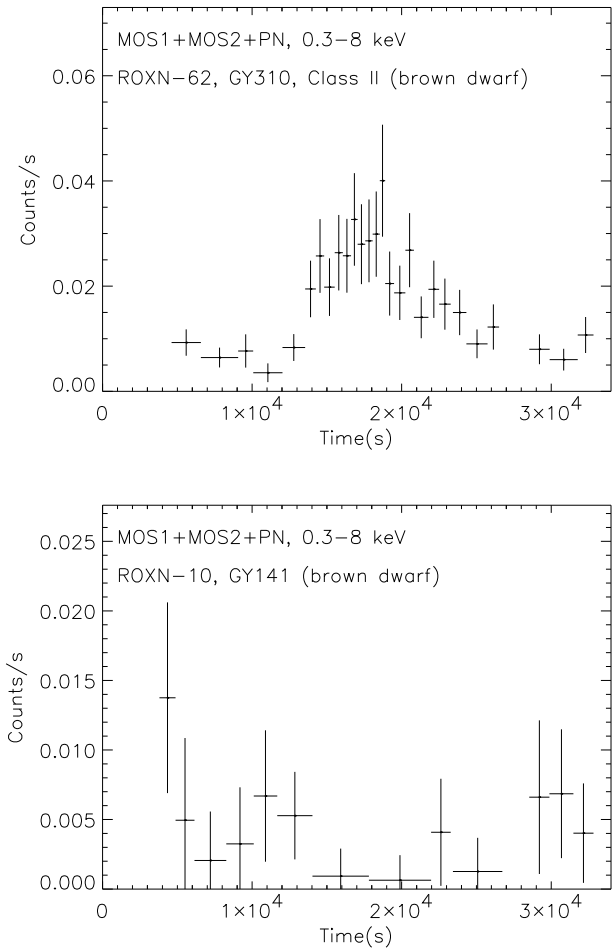}}
\caption{
X-ray background subtracted light curves of the young {\it bona fide} brown dwarfs, GY310 and GY141, obtained with {\it XMM-Newton}.
High background time intervals were suppressed (holes in the light curve). 
We use an adaptive binning to keep a constant count number in each bin (see Table 3).
}
\label{fig_lc_bd}
\end{center}
\end{figure}

A very weak X-ray emission has been reported from GY141 with {\it Chandra} (Imanishi et al.\ 2001b; source BF-S2 in Imanishi et al.\ 2003): only $\sim$8 X-ray photons were collected during the 100\,ks exposure. According to Fig.~\ref{fig_chanc} this low count rate is well below the sensitivity of our {\it XMM-Newton} observation. However we detect GY141 (ROXN-10) with a count rate of 7.7\,cts\,ks$^{-1}$, i.e., at a level $\sim$ 90 times higher than during the {\it Chandra} observation. Although we detect this brown dwarf during a phase of intense X-ray activity, we do not have enough statistics to investigate further this high X-ray state (see its background-subtracted light curve in Fig.~\ref{fig_lc_bd}).

\begin{figure}
\resizebox{\hsize}{!}{\includegraphics[angle=270]{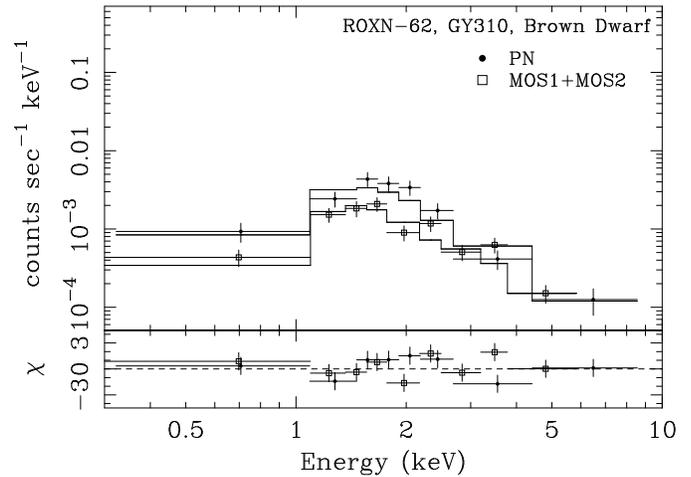}}
\caption{
X-ray spectra of the young {\it bona fide} brown dwarf GY310 during the flare state obtained with {\it XMM-Newton}. 
The filled circles and open squares indicate PN and MOS1$+$MOS2 spectra.
The solid lines show the best fit models whose spectral parameters are listed in Table 4.
}
\label{fig_spec_bd}
\end{figure}

X-ray emission was also reported from GY310 with {\it Chandra} (Imanishi et al.\ 2001a; Imanishi et al.\ 2001b; Imanishi et al.\ 2003).
The {\it Chandra} X-ray light curve shows no clear flare, but exhibits aperiodic variability by a factor 2 within the 100\,ks exposure, 
around $\sim2\times10^{-3}$\,cts\,s$^{-1}$ (Imanishi et al.\ 2001b). During our observation, GY310 (ROXN-62) displayed an X-ray flare (see Fig.~\ref{fig_lc_bd}). To our knowledge this is the first X-ray flare from a young {\it bona fide} brown dwarf with enough counts to derive its X-ray spectrum $--$ due to {\it XMM-Newton}'s large effective area. The observed count rate with MOS1+MOS2 increased from 0.007\,cts\,s$^{-1}$ to 0.04\,cts\,s$^{-1}$. The quiescent level gives an equivalent {\it Chandra} count rate of 
$2\times0.007/6.8 \sim 2\times 10^{-3}$\,cts\,s$^{-1}$, consistent with the low value previously observed by {\it Chandra}.
We obtain enough X-ray counts from GY310 during the flare to make spectral analysis presented in
 Sect.\ \ref{subsec_spec} (see Fig.~\ref{fig_spec_bd} and Table~4).
The derived column density is identical to the one found by Imanishi et al.\ (2001b). We find a somewhat higher plasma temperature in 
our flare observation (2.5\,keV, 1.7--3.6\,keV), compared to the quiescent {\it Chandra} value (1.7\,keV, 0.9--2.2\,keV), a fairly general characteristic of stellar X-ray flares.

Similar changes in the level of X-ray activity, X-ray flares, and high plasma temperatures, are ubiquitous in low-mass protostars and T Tauri stars.
This suggests that X-ray emission from brown dwarfs in their early phase of evolution are produced basically by the same solar-like, magnetic activity mechanism at work in low-mass 
protostars and T Tauri stars, and more generally in late-type stars, as also noticed in previous studies (Imanishi et al.\ 2001b, Feigelson et al.\ 2002b).

\section{X-ray properties of Young Stellar Objects in the $\rho$~Ophiuchi cloud}\label{sec_x}

\begin{figure}
\begin{center}
\resizebox{\hsize}{!}{\includegraphics[angle=270]{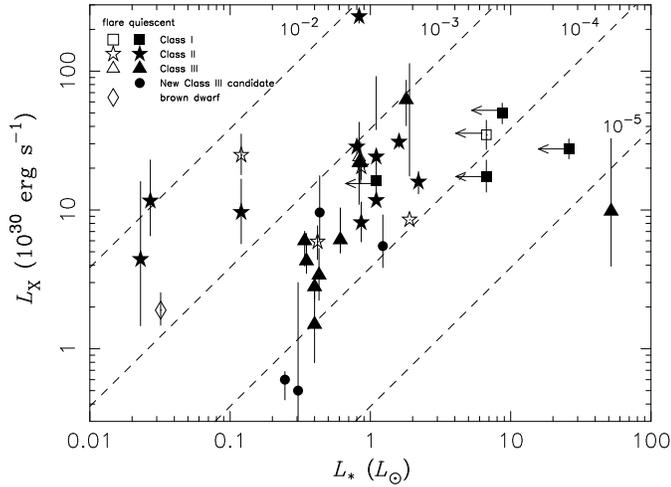}}
\caption{
X-ray luminosity, $L_X$, vs. stellar luminosity, $L_{*}$, plot for YSOs in the $\rho$ Ophiuchi cloud core F. 
The open squares, stars, triangles, and diamonds indicate Class I, II, and III sources and brown dwarfs 
in the flare state, respectively, and the filled squares, stars, triangles, and circles indicate
 Class I, II, and III sources and new Class III candidates in the quiescent state, respectively.
For Class I sources, the bolometric luminosities are plotted as the upper limits of stellar luminosities.
}
\label{lx_lbol}
\end{center}
\end{figure}

To investigate a possible correlation between the X-ray luminosity $L_X$ and the stellar luminosity $L_{*}$
 of YSOs,
 we plot $L_X$ vs. $L_{*}$ for Class I, II and III sources, new Class III candidates, and brown dwarfs in Fig.~\ref{lx_lbol}.
We calculate absorption corrected $L_X$ in the 0.5$-$10 keV band 
using the best fit model for the X-ray spectra, which are listed in Table 4.
In Fig.~\ref{lx_lbol}, we plot only the ROXN sources which are bright enough to determine their spectral parameters,
 except for ROXs20A (ROXN-27), ROXs20B (ROXN-28), WL5 (ROXN-32), WL4 (ROXN-34), and WL3 (ROXN-35), for which the individual X-ray spectra cannot be resolved.
We use $L_{*}$ from Bontemps et al.\ (2001) for Class I, II, and III sources and brown dwarfs, 
 and use our estimation of $L_{*}$ for new Class III candidates presented in section \ref{sec_ir}.
For Class I sources, as the stellar luminosity is unknown because
  the central star is invisible due to the remnant dust envelopes,
 we use the bolometric luminosity as an upper limit to the stellar luminosity.
The correlation index between $L_X$ and $L_{*}$ of Class II and III sources in the quiescent state is then $-$0.08, which indicates a weak correlation.
Indeed the $L_X / L_{*}$ ratios show a large spread, from $4.5 \times 10^{-6}$ to $1.6 \times 10^{-2}$.
New Class III candidates have similar $L_X$ and $L_{*}$ properties as other YSOs, 
which is consistent with the assumption of their YSO nature.

\begin{figure}
\begin{center}
\resizebox{\hsize}{!}{\includegraphics[angle=270]{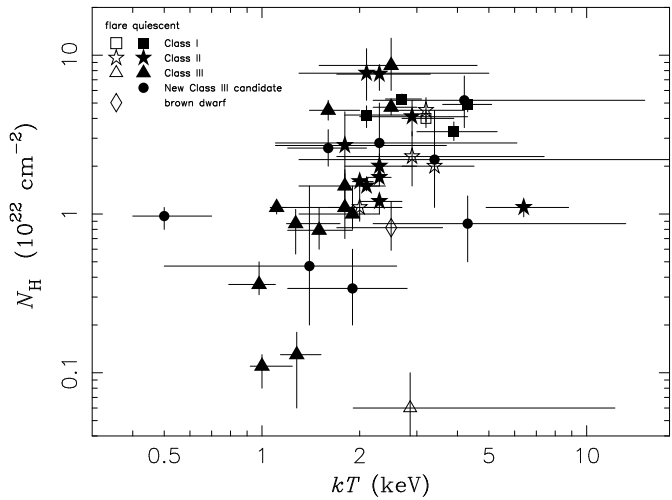}}
\caption{
$N_\mathrm{H}$ vs. $kT$ plot of the X-ray sources listed in Table 4. 
The symbols are the same as in Fig.~\ref{lx_lbol}.
For ROXN-27-28, ROXN-36, and ROXN-73 which need multi-temperature model for the spectral fitting, 
the average temperatures  weighted by the emission measures are plotted.
The errors show 90 \% confidence region.
}
\label{kt_nh}
\end{center}
\end{figure}

To compare the characteristics of the X-ray spectra of Class I, II, and III sources,
we show in Fig.~\ref{kt_nh} the scatter plot of their X-ray determined absorption column density, $N_{\rm{H}}$, vs. plasma temperature, $kT$.
For ROXN-36 (SR12A-B) and ROXN-73 (IRS55), we calculated the average temperature from the multiple components weighted by their emission measures (see above, \S 2.3).
An interesting property emerges from the $N_\mathrm{H}$ vs. $kT$ diagram: only a
few data points are seen in the upper-left region and the lower-right region.
The upper-left region is where the X-ray sources have a low plasma temperature and large absorption:
since the soft X-rays from low temperature plasmas are easily absorbed, 
 it is natural that X-rays should be hard to detect from these sources.
On the other hand, the lower-right region is where the X-ray sources have a high plasma temperature and small absorption, and are thus easy to detect, but this region contains surprisingly few data points.
This could be explained by an intrinsic effect if both the maximum temperature of YSOs and their
 circumstellar material causing absorption of X-rays decrease in the course of their evolution.
Class I sources, which are in the early stage of the evolution, are indeed located in the upper-right region of this diagram.
Class II and III sources, which are in a later stage of the evolution, are
 located in the region of lower temperature and smaller absorption than Class I sources, 
 which confirms the results obtained by {\it Chandra}  (Imanishi et al.\ 2001a).
There is also a tendency for the temperature and absorption of Class III sources to be lower than those of Class II. We conclude that, although our statistics are still limited, an evolutionary effect seems to be present from the Class I stage (high X-ray temperature, high extinction) to the Class III stage (lower X-ray temperature, lower extinction).

\section{Summary and conclusions}\label{sec_sum}

The main results of our {\it XMM-Newton} 33 ks, 30' diameter observation of the $\rho$ Ophiuchi cloud core F (i.e., the main molecular core) are as follows :

\begin{enumerate}
\item{
We detect 87 X-ray sources;
62 sources are already known from previous X-ray observations,
hence we find 25 new X-ray sources.
}
\item{ 
In the region covered by the field-of-view of both {\it XMM-Newton} and {\it Chandra} observations, 
4 X-ray sources are detected only with {\it XMM-Newton} and 38 X-ray sources are detected only with {\it Chandra}, 
 while 43 X-ray sources are detected  with both observatories. 
We have shown that these differences are explained by the difference in exposure time and sensitivity to point source detection, along with intrinsic time variability of X-ray sources (flares). 
We find that {\it Chandra} and {\it XMM-Newton} are equally sensitive for point source detection in star forming-regions for exposures of {$\sim$ 30 ks}.
}
\item{
 We detect X-ray emission from 7 Class I sources, 26 Class II sources, and 17 Class III sources.
For Class I sources, our detection rate is high (64 \%), consistent with that already observed by {\it Chandra} (Imanishi et al.\ 2001a).
}

\item{
For Class II sources, our detection rate is much lower (48 \% vs. 70 \%), which can be explained by a combination of instrumental reasons (better sensitivity for point source detection by {\it Chandra}) and intrinsic reasons (better probability to detect a source via X-ray flares in a longer exposure).
}

\item{
We find a high detection rate for Class III sources, which is not surprising in view of the fact that most Class III sources were discovered in X-rays. But we propose 15 previously unknown X-ray sources as new class III candidates, in addition to the already known 87 YSOs in the $\rho$~Oph core region (Bontemps et al.\ 2001). Pending spectroscopic confirmation, this would double the number of Class III sources known in this area.
}
\item{
We detect X-ray emission from two young bona fide brown dwarfs,  GY310 and GY141.
GY310 showed an X-ray flare during our observation, and
GY141 appeared brighter by nearly two orders of magnitude than in the {\it Chandra} observation (Imanishi et al.\ 2001a). 
Their X-ray properties suggest that the X-ray emission from young  brown dwarfs is the same as in low-mass protostars and T Tauri stars, i.e., resulting from solar-like magnetic activity, as proposed by previous studies (Imanishi et al.\ 2001b, Feigelson et al.\ 2002b).
}
\item{
We are able to extract X-ray light curves and spectra from a dozen sources, 
 and find some of them showed X-ray flares. Altogether,
from spectral fitting we confirm that there is an evolutionary trend, from Class I sources showing higher temperature and larger
  absorption, to Class II and III sources showing lower temperature and smaller extinction, as previously reported by the {\it Chandra} observation (Imanishi et al.\ 2001a).
}
\end{enumerate}

\begin{acknowledgements} 
H.O. acknowledges the Postdoctoral Fellowship for Research Abroad
 of the Japan Society for the Promotion of Science, and Conseil national des astronomes et physiciens.
\end{acknowledgements}

\appendix
\section{Method to compute the X-ray count rate upper limit }
\subsection{XMM-Newton data}

For each instrument, MOS1, MOS2, and PN, we compute with the SAS command {\tt eexpmap} the 
corresponding exposure map with a spatial resolution of 
2$^{\prime\prime}$
for X-ray photons of 1.5\,keV energy, and normalize it by its maximum value located on-axis. 
The resulting bidimensionnal array, $\cal E$, represents the spatial effective area 
variations, including CCD gaps, bad pixels, and vignetting, relative to the on-axis value. 
From $\cal E$ we define an exposure mask, $\cal M$, 
having  $\cal M$(i)=0 if $\cal E$(i)=0, and $\cal M$(i)=1 if $\cal E$(i)$\neq$0. 
This exposure mask shows only CCD gaps.

We estimate for each position of undetected Chandra source the local average background. 
We extract the event number, $N_\mathrm{bgd}$, inside the 15\arcsec-radius disk, 
$\cal D_\mathrm{bgd}$, of geometrical area $A_\mathrm{bgd}^\mathrm{geo}$ centered 
on the Chandra source position. 
This extraction region contains $n^{\cal E}_\mathrm{bgd}$ sky pixels of $\cal E$, 
and $n^{*{\cal M}}_\mathrm{bgd}$ sky pixels of $\cal M$ where $\cal M$ is equal to 1, 
i.e. which are not CCD gaps. The net pixel area of $\cal D_\mathrm{bgd}$ is~: 
\begin{equation}
A_\mathrm{bgd}^\mathrm{net} = A_\mathrm{bgd}^\mathrm{geo} \, n^{*{\cal M}}_\mathrm{bgd}\,/\,n^{\cal E}_\mathrm{bgd}.
\end{equation}
The average local background per sky pixel is obtained by~:
\begin{equation}
<\!B\!>\,= N_\mathrm{bgd}/A_\mathrm{bgd}^\mathrm{net}. 
\end{equation}

This 15\arcsec-radius size is a compromise for crowded region. 
With this local method, local background variations as PSF wings of neighbouring X-ray source, 
or trails due to out-of-time events of bright sources, are 
automatically included in our background estimate. This would not be the case 
if we had used the spline-smoothed background built by {\tt esplinemap}, because 
it uses a `Swiss-cheese' method which subtracts the source PSF from the X-ray image. 

Then we compute the PSF radius encircling $f_\mathrm{ee}=50\%$ of the PSF 
energy, $R(f_\mathrm{ee})$, using the radial average of the calibration images in the CCF.
We extract the event number, $N_\mathrm{src+bgd}(f_\mathrm{ee})$, within 
the $R(f_\mathrm{ee})$-radius disk, $\cal D_\mathrm{src}$, of geometrical area 
$A_\mathrm{src}^\mathrm{geo}$ centered on the Chandra source position. 
These counts are the sum of background and source 
counts, $N_\mathrm{src}(f_\mathrm{ee})$ and $N_\mathrm{bgd}(f_\mathrm{ee})$, 
respectively. 
The extraction region $A_\mathrm{src}^\mathrm{geo}$ contains $n^{\cal E}_\mathrm{src}$ sky pixels of $\cal E$, 
and $n^{*{\cal M}}_\mathrm{src}$ sky pixels of $\cal M$ where $\cal M$ is equal to 1. 
The net pixel area of $\cal D_\mathrm{src}$ is~: 
\begin{equation}
A_\mathrm{src}^\mathrm{net} = A_\mathrm{src}^\mathrm{geo} \, n^{*{\cal M}}_\mathrm{src} \,/\, n^{\cal E}_\mathrm{src}.
\end{equation}
We compute for $N_\mathrm{src+bgd}(f_\mathrm{ee})$ its 99.9989\% confidence level (corresponding 
to 4.4$\sigma$ in Gaussian statistics, i.e. the threshold of our detection algorithm) upper limit, 
$N^\mathrm{ul}_\mathrm{src+bgd}(f_\mathrm{ee})$, using the approximate 
formula (10) of Gehrels (1986).
Assuming that the background is constant on $\cal D_\mathrm{src}$, the upper limit on the source count 
in $\cal D_\mathrm{src}$ is~:
\begin{eqnarray}
N^\mathrm{ul}_\mathrm{src}(f_\mathrm{ee}) & = & \left\{ N_\mathrm{src+bgd}^\mathrm{ul}(f_\mathrm{ee}) 
- <\!B\!>  A_\mathrm{src}^\mathrm{net} \right\} \nonumber \\
& & \times \frac{n^{\cal E}_\mathrm{src}}{\sum_{\mathrm{i} \in {\cal D_\mathrm{src}}} {\cal E}(i)}\,.
\end{eqnarray}
The right hand side multiplicative factor corrects in average from loss of sensitivity due to bad pixels, CCD gaps, and vignetting. 
For Chandra sources which are over CCD gaps or bad pixels of XMM-Newton (a minority in our sample), 
a more accurate correction would involve a convolution of the PSF shape with the exposure mask. 

The upper limit on the total source count is hence :
\begin{equation}
N^\mathrm{ul}_\mathrm{src}=N^\mathrm{ul}_\mathrm{src}(f_\mathrm{ee})\,/\,f_\mathrm{ee} \,. 
\end{equation}
Finally, the count rate upper limit is obtained by dividing $N^\mathrm{ul}_\mathrm{src}$ by the 
livetime of the corresponding CCD.

We compute the count rate upper limit for the sum of the MOS1, MOS2, and PN data, the so-called 
EPIC instrument, $CR^\mathrm{ul}_\mathrm{src,EPIC}$, using this set of straightforward formulae using 
the previous notation, where $i \in [\mathrm{MOS1,MOS2,PN}]$~:

\begin{equation}
N_\mathrm{src+bgd,EPIC}(f_\mathrm{ee})=\sum_\mathrm{i}\,N_\mathrm{src+bgd,i}(f_\mathrm{ee})\, ,
\end{equation}

\begin{eqnarray}
N^\mathrm{ul}_\mathrm{src,EPIC}(f_\mathrm{ee}) & = &  \left\{ N_\mathrm{src+bgd,i}^\mathrm{ul}(f_\mathrm{ee}) -  \sum_\mathrm{i} <\!B\!>_\mathrm{i} A_\mathrm{src,i}^\mathrm{net} \right\} \nonumber \\
 & &  \times \frac{\sum_\mathrm{i} n^{{\cal E}_\mathrm{i}}_\mathrm{src} }{\sum_\mathrm{i} \sum_{\mathrm{j} \in {\cal D}_\mathrm{src,i}} {\cal E}_\mathrm{i}(j)}
\, ,
\end{eqnarray}

\begin{equation}
N^\mathrm{ul}_\mathrm{src,EPIC} = N^\mathrm{ul}_\mathrm{src,EPIC}(f_\mathrm{ee})\,/\,f_\mathrm{ee} \, ,
\end{equation}

\begin{equation}
CR^\mathrm{ul}_\mathrm{src,EPIC}=N^\mathrm{ul}_\mathrm{src,EPIC}\,/\/\sum_\mathrm{i} LIVETIME_\mathrm{\,i}\, .
\end{equation}

\subsection{Chandra data}

We get the level 2 data of the Core F observation of the $\rho$ Ophiuchi dark cloud 
(sequence number 200060, observation ID 635) from the Chandra data 
archive\footnote{{\tt http://asc.harvard.edu/cgi-gen/cda/retrieve5.pl\,.}}. 
We compute the exposure map with a resolution of 4$^{\prime\prime}$ for X-ray photons of 1.5\,keV energy,
 and normalize it by its maximum value (on-axis), 
from which we derive an exposure mask.
We proceed as for XMM-Newton data, using for the PSF radius encircling 50\% of the PSF 
energy the formula given by Feigelson et al.\ (2002b)~: 
$R(50\%)=0.43-0.1\,\theta +0.05\,\theta\,^2$, where $\theta$ is the off-axis angle in arcmin.

We note that for Chandra data, thanks to the large satellite wobbling, bad pixels and CCD gaps 
are smooth in the exposure map, they do not produce holes with sharp edges, hence with the above notation we have 
$n^{*{\cal M}} / n^{\cal E} =1$.
By combining (1)--(3), we find a formula identical to the formula (6) of Feigelson et al. (2002b) 
when replacing the 90\% upper limit count by the observed count.

\section{Comparison between {\it Chandra} and {\it XMM-Newton} determination of $N_\mathrm{H}$}

\begin{figure}
\resizebox{\hsize}{!}{\includegraphics[angle=270]{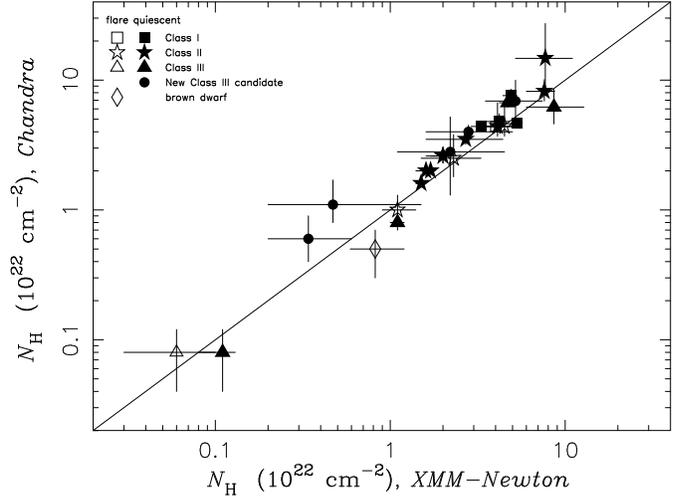}}
\caption{
 Comparison of $N_\mathrm{H}$ determinations from spectral fitting between {\it Chandra} and {\it XMM-Newton}.
We plot for comparison the diagonal with a slope equal to 1.
The symbols are the same as in Fig. 10. The errors show 90 \% confidence region.
{\it Chandra} and {\it XMM-Newton} estimates of $N_\mathrm{H}$ are consistent taking account the error bars.
}
\label{fig_nh}
\end{figure}

To check the reliability of the X-ray derived values of $N_\mathrm{H}$,
 we make a plot of the $N_\mathrm{H}$ values determined by {\it Chandra} vs. those determined by {\it XMM-Newton} for
 the X-ray sources with spectral fitting data (see Table.\ 4) in the overlapping field-of-views of the two satellites (see Fig.\ \ref{fig_nh}).
A best fit value of the ratio of the $N_\mathrm{H}$ values from {\it Chandra} and {\it XMM-Newton} obtained from a linear fit is 0.96 ( 0.93$-$0.99, 90 \% confidence region), 
 indicating that the both determinations of $N_\mathrm{H}$ values are consistent better than 10\%.  

Uncertainties of the X-ray derived $N_\mathrm{H}$ values were discussed for both {\it Chandra} and {\it XMM-Newton} in Vuong et al. (2003).
They showed that the use of the recently revised solar abundances (Holweger 2001, Allend Prieto et al. 2001, 2002) in X-ray spectral fitting increases $N_\mathrm{H}$ values by $\sim$20 \%.
  Both in the X-ray spectral fitting by {\it Chandra} (Imanishi et al.\ 2001a) and by {\it XMM-Newton} (this work, see Sect.\ \ref{subsec_spec}), the $N_\mathrm{H}$ values were determined using WABS absorption model in XSPEC where solar standard metal abundances are used.
If the abundance effect shown by Vuong et al. (2003) is taken into account, all the $N_\mathrm{H}$ values derived by {\it Chandra} and {\it XMM-Newton} would increase and all the data points in Fig. \ref{fig_nh} would move towards upper-right direction by $\sim$20 \% along the diagonal.

\end{document}